\documentclass[%
 aip,
 amsmath,amssymb,
 reprint,%
]{revtex4-1}
\usepackage{mathtools}
    \usepackage{booktabs}
    \usepackage{graphicx}
    \usepackage{xcolor}
    \usepackage{dcolumn}
    \usepackage{bm}
    \usepackage{hyperref}
    \DeclareMathAlphabet{\mathpzc}{OT1}{pzc}{m}{it}
    
    \usepackage[utf8]{inputenc}
    \usepackage[T1]{fontenc}
    \usepackage{mathptmx}
 \setcitestyle{super}

\newcommand*{\citen}[1]{%
  \begingroup
    \romannumeral-`\x 
    \setcitestyle{numbers}%
    \cite{#1}%
  \endgroup   
}

\newcommand{\norm}[1]{\left\lVert#1\right\rVert}

\begin{document}
    
    
\title{FCHL revisited: faster and more accurate quantum machine learning}
    \author{Anders S. Christensen}
    \affiliation{Institute of Physical Chemistry,
    National Center for Computational Design and
    Discovery of Novel Materials (MARVEL),
    Department of Chemistry,
    University of Basel,
    Klingelbergstrasse 80,
    CH-4056 Basel, Switzerland}
    \author{Lars A. Bratholm}
    \affiliation{School of Mathematics, University of Bristol, Bristol, BS8 1TW, UK}
    \affiliation{School of Chemistry, University of Bristol, Bristol BS8 1TS, UK.}
    \author{Felix A. Faber}
    \affiliation{Institute of Physical Chemistry,
    National Center for Computational Design and
    Discovery of Novel Materials (MARVEL),
    Department of Chemistry,
    University of Basel,
    Klingelbergstrasse 80,
    CH-4056 Basel, Switzerland}    
    \author{O. Anatole von Lilienfeld}%
    \email{anatole.vonlilienfeld@unibas.ch}
    \affiliation{Institute of Physical Chemistry,
    National Center for Computational Design and
    Discovery of Novel Materials (MARVEL),
    Department of Chemistry,
    University of Basel,
    Klingelbergstrasse 80,
    CH-4056 Basel, Switzerland}
    
    \date{\today}

    \begin{abstract}

    We introduce the FCHL19 representation for atomic environments in molecules or condensed-phase systems.
    Machine learning models based on FCHL19 are able to yield predictions of atomic forces and energies of query compounds with chemical accuracy on the scale of milliseconds.
    FCHL19 is a revision of our previous work\cite{faber2017alchemical} where the representation is discretized and the individual features are rigorously optimized using Monte Carlo optimization.
    %
    %
    Combined with a Gaussian kernel function that incorporates elemental screening, chemical accuracy is reached for energy learning on the QM7b and QM9 datasets after training for minutes and hours, respectively.
    %
    %
    %
    The model also shows good performance for non-bonded interactions in the condensed phase for a set of water clusters with an MAE binding energy error of less than 0.1 kcal/mol/molecule after training on 3,200 samples.
    For force learning on the MD17 dataset, our optimized model similarly displays state-of-the-art accuracy with a regressor based on Gaussian process regression.
    When the revised FCHL19 representation is combined with the operator quantum machine learning regressor, forces and energies can be predicted in only a few milliseconds per atom.
    The model presented herein is fast and lightweight enough for use in general chemistry problems as well as molecular dynamics simulations.
    %
    
    \end{abstract}
    
    \maketitle
    
    \section{Introduction}
    Approximate models have been used to make predictions in chemistry since the beginning of theoretical chemistry. 
    In recent years, however, data-driven machine learning (ML) models which can make predictions across chemical space with chemical accuracy are becoming increasingly common in literature.
    
    Tasks such as molecular dynamics (MD) simulations and geometry optimizations have been a standard tool in the toolbox of the computational chemist for many years, and several machine learning models now provide the gradients necessary to carry out such tasks.\cite{GAPtutorial,Neuralnetworks_Scheffler2004,Neuralnetworks_BehlerParrinello2007,BehlerPerspective2016,ANI_IsayevRoitberg2017,smith2017ani,Botu2015,RampiMLQMMM,Botu2016,Huan2017,Zhenwei2015,gubaev2018machine,SNAP_Aidan2015,CovariantKernelsSandro2016,Glielmo2018,schutt2018schnet,schutt2019schnetpack,grisafi2018symmetry,DeepMDZhang2018,UnkePhysNet2019}
    We have previously published a machine learning model based on the Faber-Christensen-Huang-Lilienfeld (FCHL18) representation which performs very well on chemical compounds across chemical space,\cite{faber2017alchemical} as well as a proof-of-concept implementation of learning and prediction of response properties based on this model,\cite{christensen2018operators} such as atomic force, normal modes, dipole moments, and even IR spectra.

    While our FCHL18-based models yielded state-of-the-art accuracy on several benchmark sets,\cite{faber2017alchemical,christensen2018operators} the applicability was in some cases hindered by poor computational performance, and proper hyperparameter optimization of the model has been computationally unfeasible.
    Whereas FCHL18 solves an analytical integral to compare atomic environments in order to learn properties of chemical compounds, other ML models use discretized representations that can be handled with far greater computational efficiency.\cite{rogers2010extended,RuppPRL2012,AssessmentMLJCTC2013,constsize2018,FourierDescriptor,chmiela2017machine,Chmiela2019sGDML,Neuralnetworks_Behler2011,ANI_IsayevRoitberg2017,Gastegger2018wacsf,SOAPDiscretized2019}

    In this work, we present a discretized representation for chemical compounds based on our earlier work in Ref.~\citen{faber2017alchemical}, which allows for a very fast evaluation of the L2-distance between two representations.
    A rigorous Monte Carlo optimization of the model parameters is performed in order to find a set of universally transferable hyperparameters that yield ML models of high accuracy without any need for re-optimization.
    We include a detailed review of different kernel-based models with which the representation can be used and highlight their strengths, differences, and shortcomings.
    Lastly, we thoroughly benchmark the predictive accuracy of our models on several established datasets of chemical compounds from the literature.
    In addition to benchmarking the accuracy of energy and force prediction, we also present timings of our model in order to demonstrate the applicability.

    
    \section{Theory}
    This section first introduces the representation used to describe atomic environments throughout this work.
    Secondly, a number of kernel-based machine learning methods which can be used with the representation are discussed.
    While the representation could in principle also be used favorably with feed-forward neural networks, this paper focuses solely on kernel-based methods.

    \subsection{Representation}
    We have previously compared ML models based on a number of different representations for the QM9 dataset.\cite{googlePaper2017,faber2017alchemical}
    Based on these studies it is apparent that the currently best-performing representations contain certain similarities, although the exact implementations differ.
    Some of the best-performing representations for kernel-based machine learning for chemical compounds are the smooth overlap of atomic densities (SOAP),\cite{Sandip2016,BartokGabor_Descriptors2013} spectrum of London and Axilrod-Teller-Muto (SLATM),\cite{Bing2016} the many-body descriptor of of Pronobis \textit{et al.},\cite{pronobis2018many} and FCHL18 representations,\cite{faber2017alchemical} while variants of the atom-centered symmetry functions (ACSF) of Behler\cite{Neuralnetworks_Behler2011,ANI_IsayevRoitberg2017,Gastegger2018wacsf} have been shown to perform well for feed-forward neural networks.
    In brief, these methods contain some terms that are similar:
    1) a two-body term that relates to the radial distribution between a central atom and other nearby atoms in its local environment and 2) a three-body term that similarly relates to, for example, distribution of angles and/or distances between atoms in the local environment of the atom.

    In this paper we construct a new atom-centered representation termed FCHL19 that contains such two-and three-body terms and demonstrate that this leads to similar performance.
    The FCHL19 representation is based on the FCHL18 representation,\cite{faber2017alchemical} but is discretized in a manner very similar to the well-known ACSF of Behler.\cite{Neuralnetworks_Behler2011}

    In order to enable faster and more memory efficient machine learning models, it is key that the input representation is as small as possible compared to the information it holds, as evaluation and training times scales linearly/quadratically with representation size.
    %
    %
    
    We show that when the parameters of our new representation are optimized properly, the result is a representation that is compact in size---ensuring faster machine learning algorithms---without loss in predictive accuracy. 

    Briefly described, the representation is a vector that encodes the atomic environment of an atom in a chemical compound.
    It consists of a two-body term which encodes radial distributions between the central atoms and neighboring atoms of a given element type.
    Additionally, the representation contains a three-body term that encodes the mean distances and angles between the atom and neighboring pairs of atoms of given element types.
    %
    %
    
    The representation does not contain an explicit one-body term and, for performance reasons, we do not consider terms of higher order than three-body, but it is possible that the inclusion of such terms could lead to even higher predictive accuracy.\cite{googlePaper2017}
    
    The two- and three-body components of the representation are described in detail in the following text.
    The procedure to obtain the hyperparameters of the representation is detailed in the "Methodology" section \ref{sec:optparams}, while the optimized parameters are presented in Table \ref{tab:parameters} in Appendix A.
    
    \subsubsection{Two-body function}
    \begin{figure}
        \centering
        \includegraphics[width=\linewidth]{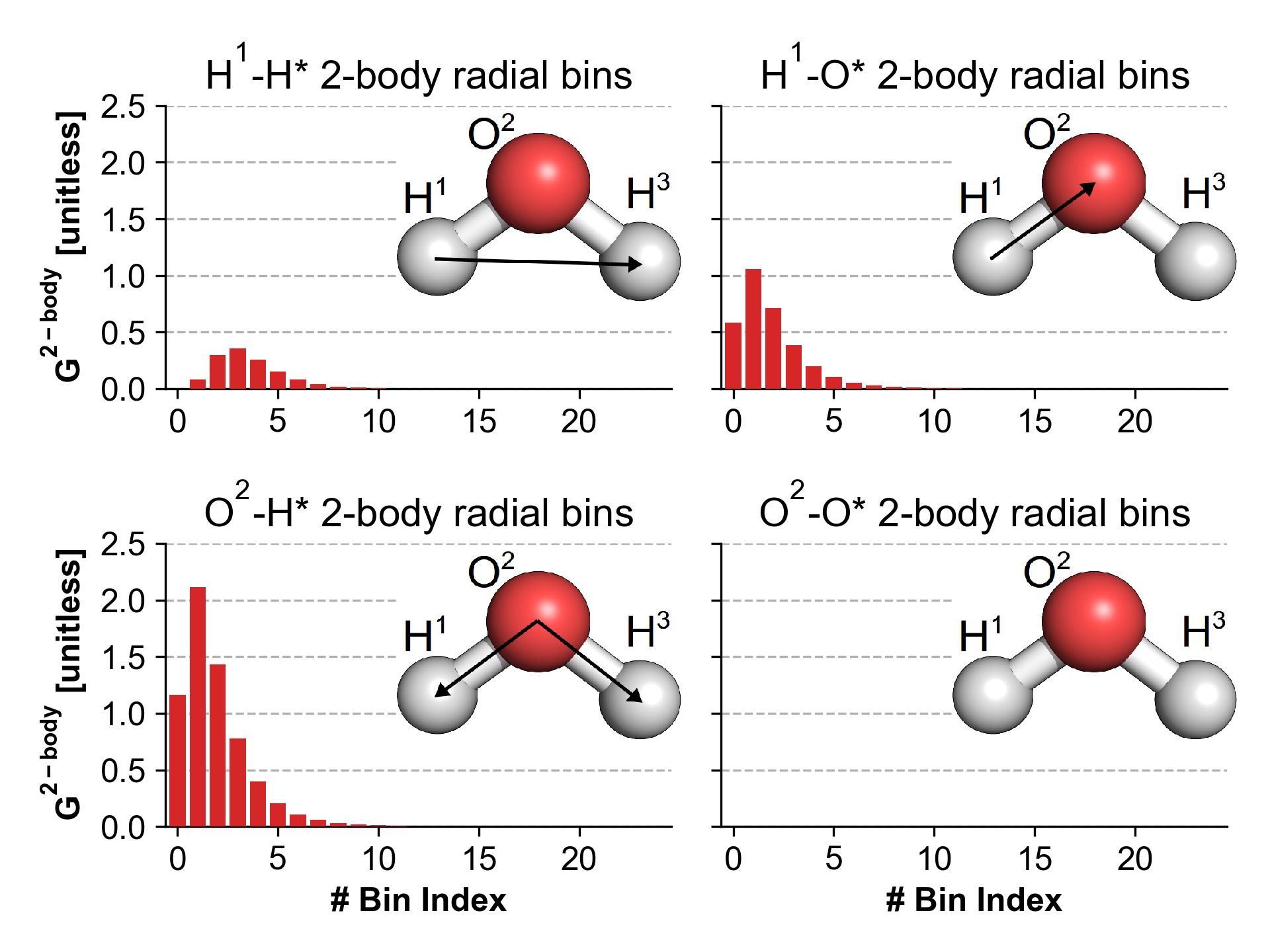}
        \caption{\label{fig:twobody}The values of four unique types of two-body radial basis functions in a water molecule are displayed. The radial spectrum is divided into 24 bins, with $r_\text{cut} = 8.0$ \AA, $w = 0.32$ \AA~and $N_2 = 1.8$.~The top row contains the radial basis functions for the first H atom and the bottom row for the oxygen atom.
        The distances that are used to produce the basis functions in each spectrum are marked with black arrows.}
    \end{figure}
    
    For a given central atom, a set of radial basis functions is constructed for each unique type of element in the data set.
    Each of the $n_{R_{s_2}}$ basis functions in this set is placed on an equidistant grid from $\tfrac{r_{cut}}{n_{R_{s_2}}}$ to $r_{cut}$, with $r_{cut}$ being the cutoff radius.
    We found it advantageous to use log-normal distribution functions for the radial functions, compared to Gaussian functions as used in our previous work.\cite{faber2017alchemical}
    We note that this is an empirical choice and it is possible that a better distribution function could be found, for example from using an optimization procedure.
    %
    %
    The log-normal radial basis functions take the form:
    \begin{equation}
    G^\text{2-body} =  \xi_2\left(r_{IJ}\right)  
    f_\text{cut}\left(r_{IJ}\right) \tfrac{1}{R_s\sigma\left(r_{ij}\right)\sqrt{2\pi}} 
    \exp\left(- \frac{\left( \ln R_s - \mu\left(r_{ij}\right) \right)^2}{2\sigma\left(r_{ij}\right)^2}\right)
    \end{equation}
    where $R_s$ is the distance location of the grid point, and $\mu\left(r_{ij}\right)$ and $\sigma\left(r_{ij}\right)$ are parameters of the log-normal distribution, which in turn depend on the interatomic distance, $r_{IJ}$,  and a hyperparameter, $w$, given as follows:
    \begin{equation}
    \mu\left(r_{ij}\right) = \ln \left( \frac{r_{IJ}}{\sqrt{1+\frac{w}{r_{IJ}^2}}}\right)
    \quad \text{and} \quad
    \sigma\left(r_{ij}\right)^2 =  \ln\left( 1 + \frac{w}{r_{IJ}^2}\right)
    \end{equation}
    The two-body scaling function, $\xi_2\left(r_{IJ}\right)$, serves the purpose of applying a higher regression weight to terms that are more likely to contribute substantially to the total energy, thus increasing the accuracy of the machine learning procedure for properties that relate to the total energy.
    Similarly to previous studies,\cite{amons2017,faber2017alchemical} we found the following form to be suitable:
    \begin{equation}
    \xi_2\left(r_{IJ}\right) = \frac{1}{r_{IJ}^{N_2}}
    \end{equation}
    where the exponent $N_2$ is hyperparameter of the representation.
    Finally, the soft cut-off function used here is:
    \begin{equation}
    f_\text{cut}\left(r_{IJ}\right) =
    \begin{cases} \frac{1}{2}\left( \cos\left(\frac{\pi~ r_{IJ}}{r_\text{cut}}\right)+1\right) &  \text{if} \quad r_{IJ} \leq r_\text{cut}\\
    0 & \text{if} \quad r_{IJ} > r_\text{cut}\\
    \end{cases}
    \end{equation}
    Thus, the hyperparameters of the two-body term are the width parameter of the log-normal distributions, $w$, the exponent of the scaling function, $N_2$, the cut-off distance, $r_\text{cut}$ and the number of radial basis functions $n_{R_{s_2}}$.
    Optimized values of these parameters are presented in Table \ref{tab:parameters} in Appendix A.

    A graphical representation of the two-body function for an H and an O atom in a water molecule is displayed in Fig.~\ref{fig:twobody}.
    For each atomic environment in the water molecule, the minimal representation will contain two radial distributions, x-H and x-O.
    Thus, the size of the two-body term scales linearly with the number of possible elements in the atomic environment.

    \subsubsection{Three-body function}
        
    \begin{figure}
        \centering
        \includegraphics[width=\linewidth]{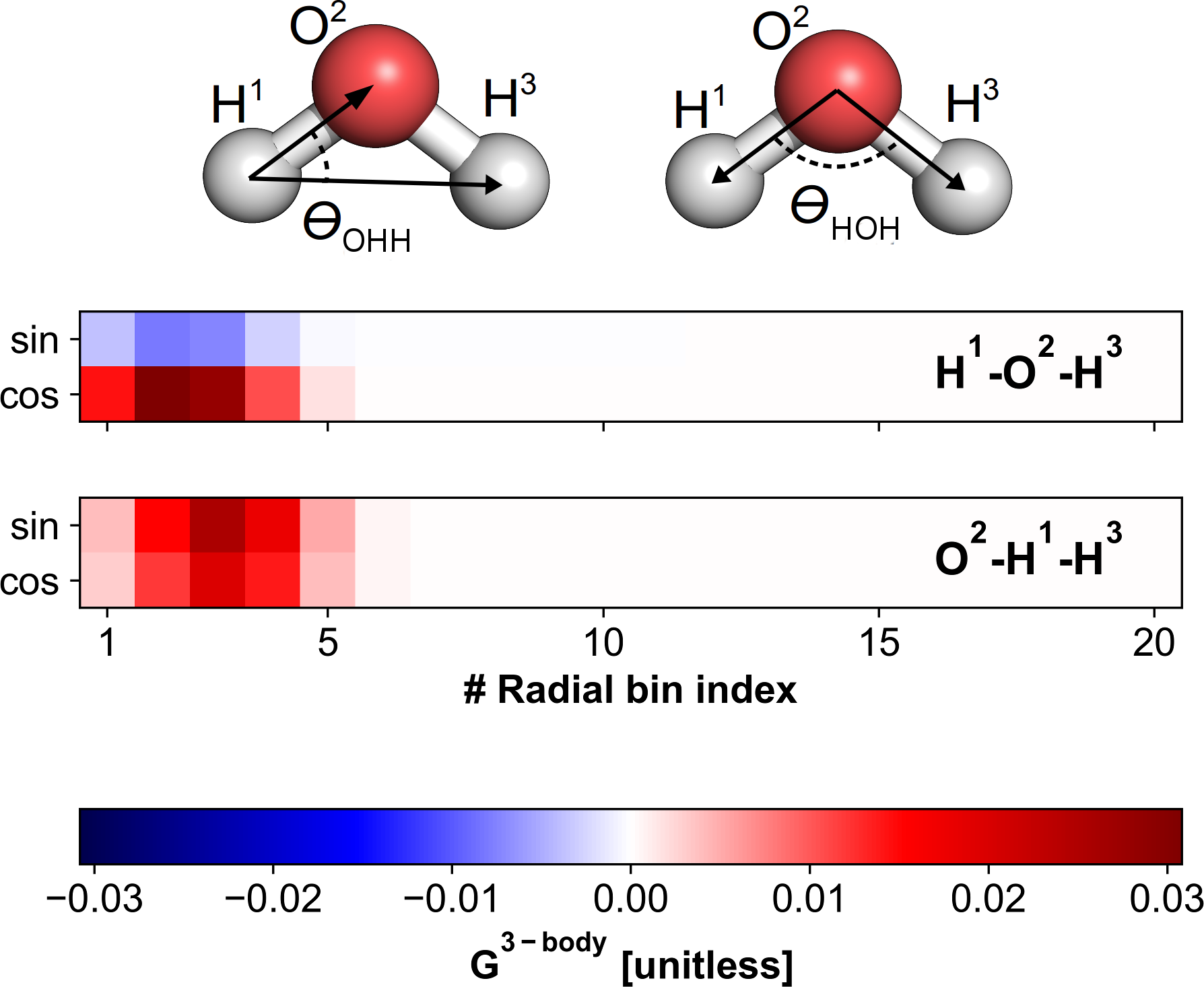}
        \caption{\label{fig:threebody}  The three-body basis functions are plotted for the two unique three-body terms in the water molecule, corresponding to the O$^2$-H$^1$-H$^3$ and H$^1$-O$^2$-H$^3$ angles displayed at the top. The atoms are numbered for clarity.
            }
    \end{figure}
    
    The three-body function encodes the distances from an atom to neighboring pairs of atoms in the environment of the atom, as well as the angle between the triplet, and the element types of the neighbors.
    The resulting function is a product of the following terms:
    \begin{equation}
    G^\text{3-body} =  \xi_3 G^\text{3-body}_\text{Radial} G^\text{3-body}_\text{Angular}f_\text{cut}\left(r_{IJ}\right) f_\text{cut}\left(r_{JK}\right) f_\text{cut}\left(r_{KI}\right)
    \end{equation}
    The radial part is similar to the radial part in the ACSFs used in the ANI-1 neural network:\cite{Neuralnetworks_Behler2011,smith2017ani}
    \begin{equation}
    G^\text{3-body}_\text{Radial} = \sqrt{\frac{\eta_3}{\pi}}\exp\left( -\eta_3 \left(\tfrac{1}{2}\left(r_{IJ} + r_{IK}\right) - R_s \right)^2\right)
    \end{equation}
    where $\eta_3$ is a parameter that controls the width of the radial distribution functions and again $R_s$ is the location of the radial gridpoints.
    Finally, the three-body scaling function, $\xi_3$ is the Axilrod-Teller-Muto term\cite{AxilrodTeller,Muto1943} with modified exponents:\cite{faber2017alchemical,amons2017}
    \begin{equation}
    \xi_3 = c_3 \frac{1 + 3\cos\left(\theta_{KIJ}\right)\cos\left(\theta_{IJK}\right)\cos\left(\theta_{JKI}\right) }{\left(r_{IK}r_{JK}r_{KI}\right)^{N_3}}
    \end{equation}
    Here $\theta_{KIJ}$ is the angle between the three atoms $K$, $I$, and $J$ and $c_3$ is a weight term that balances the weight of the three-body part relative to the two-body part.

    The angular term is similar to  the Fourier-series expansion previously introduced in Ref.~\citen{faber2017alchemical}:
    \begin{align}
        G_n^{\mathrm{cos}}&=\exp\left(-\frac{\left(\zeta  n\right)^2}{2}\right) \left(\cos{\left(n \theta_{KIJ}\right)} - \cos\left(n\left(\theta_{KIJ}+\pi\right)\right)\right)\\
        G_n^{\mathrm{sin}}&= \exp\left(-\frac{\left(\zeta  n\right)^2}{2}\right)\left(\sin{\left(n \theta_{KIJ}\right)} - \sin\left(n\left(\theta_{KIJ}+\pi\right)\right)\right)
    \end{align}
    where $\zeta$ is a hyperparameter describing the width of the angular Gaussian function and $n>0$ is the expansion order.
    With a sufficiently large value of $\eta_3$, the angular spectrum can in many cases be almost completely recovered with only the first Fourier terms.\cite{faber2017alchemical}
    This is in part due to the fact that there is only room for a limited number of atoms in the local environment at a certain distance, and the angular spectra are therefore rarely very crowded for short distances.
    In the rest of this work, only the two $n=1$ cosine and sine terms are used, i.e.~$G^\text{3-body}_\text{Angular} \in \{ G_1^{\mathrm{cos}}, G_1^{\mathrm{sin}}\}$.

    Since the number of the three-body functions scales as $\mathcal{O}\left( N^2\right)$ with the number of possible different elements in the chemical compounds, they comprise a much larger part of the representation than the two-body part.
    A graphical representation of the three-body terms for the atomic environments in a water molecule is displayed in Fig.~\ref{fig:threebody}.

    %
    \subsection{Machine learning}
    
    In the following subsections, we discuss four kernel-based regressors that are also used in this study.
    First, the kernel ridge regression (KRR) method to learn the energy of chemical compounds is discussed.
    Next, three different regressors to learn forces and energies of chemical compounds are reviewed, namely "operator quantum machine learning" (OQML),\cite{christensen2018operators} Gaussian process regression (GPR),\cite{RasmussenWilliams,sonjamathias} and finally "gradient-domain machine learning" (GDML).\cite{chmiela2017machine,Chmiela2019sGDML}

    In this section, lower-case indices denote the index of a chemical compound, while upper-case indices denotes the index of the atomic centers in the chemical compound, and finally asterisks are used to denote relation to a query compound or query atomic center.

    \subsubsection{Kernel Ridge Regression (KRR)}
    It is well-established that KRR---despite its simplicity---is one of the most powerful methods to learn energies of chemical compounds.\cite{RuppPRL2012,BobPaper,BartokGabor_Descriptors2013,AssessmentMLJCTC2013,Bing2016,Sandip2016,googlePaper2017,faber2017alchemical}
    %
    In KRR the energy, $U^{*}$, of a query compound, $c$, can be decomposed into the sum of atomic energies. 
    These are calculated in a basis of kernel functions placed on the atoms of the chemical compounds in the training set.
    That is:
    \begin{align}
    	U^{*}_{c} = \sum_{I \in c} U^{*}_\text{local}\left(\mathbf{q}^{*}_I\right) 
    	=\sum_{I \in c}  \sum_{j} \sum_{J \in j} \mathpzc{k}\left( \mathbf{q}_J, \mathbf{q}_{I}^{*}\right) \alpha_j\label{eq:local_decomposition}
    \end{align}
    where $I$ and $J$ are atoms in the query and training compounds $c$ and $j$, respectively.
    $\mathbf{q}_J$, $\mathbf{q}_{I}^{*}$ are their representation and $\alpha_j$ is the $j$'th regression coefficient.
    This can be written in matrix notation:
    \begin{equation}
            \mathbf{U} = \mathbf{K}^\text{KRR} \bm{\alpha}^\text{KRR}\label{eq:krr}
    \end{equation}
    where the elements of the KRR kernel matrix are given by the sums over the pair-wise kernels between the atoms in two compounds, 
    \begin{equation}
            \mathbf{K}^\text{KRR}_{ij} = \sum_{I \in i}  \sum_{J \in j} \mathpzc{k}\left( \mathbf{q}_J, \mathbf{q}_{I}^{*}\right)
    \end{equation}
    and $ \bm{\alpha}^\text{KRR}$ is the regression coefficient vector.

    These regression coefficients can be obtained by fitting Eq.~\ref{eq:krr} to the energies of a training set in the basis of the same set of compounds.
    In KRR this is done by minimizing the following cost function:
    \begin{equation} \label{eq:krr_lagrangian}
        J(\bm{\alpha}^\text{KRR}) = \tfrac{1}{2}\| \mathbf{K}^\text{KRR} \bm{\alpha}^\text{KRR} - \mathbf{U} \|_2^2 
            + \tfrac{\lambda}{2} \left(\bm{\alpha}^\text{KRR}\right)^T \mathbf{K}^\text{KRR} \bm{\alpha}^\text{KRR} 
    \end{equation}
    which has the following closed-form solution:
    \begin{equation}
             \bm{\alpha}^\text{KRR}  =  \left( \mathbf{K}^\text{KRR} + \mathbf{I} \lambda \right)^{-1} \mathbf{U}.
    \end{equation}
    $\lambda$ is a typically small number which is added to the diagonal of the kernel matrix in order to regularize and ensure numerical stability when the the kernel is inverted.\cite{tikhonov1977} 

    We have previously shown how KRR with FCHL18 yields systematically improving property predictions that reach state-of-the art accuracy for many system classes including molecules and materials.\cite{faber2017alchemical}
    %
    %
    %
    %
    
    \subsubsection{Operator Quantum Machine Learning (OQML)}
    It is advantageous to also include forces in the training step if available, as this both improves energy and force prediction.
    In the operator quantum machine learning (OQML) approach introduced in ref.~\citen{christensen2018operators}, the model is trained on the energy and forces simultaneously.

    The kernel is expanded in a basis of kernel functions placed on the atomic environments of each atom in the training set.
    Effectively, this extends the number of regression coefficients to the total number of atoms in the training set rather than the number of chemical compounds as for KRR.
    In the following we refer to this non-square kernel as  $\mathbf{K}^\text{OQML}$.

    In addition to the energies, $\mathbf{U}$, it is possible to include the forces, $\mathbf{F}$, in the training step by applying the force operator to the kernel and solving the regression coefficients for both the energy and forces simultaneously.
    The equation that is solved during the training step is
    \begin{equation}\label{eq:oqml}
        \begin{bmatrix}
            \mathbf{U} \\
            \mathbf{F} 
        \end{bmatrix} = \begin{bmatrix}
            \mathbf{K}^\text{OQML} \\
            -\frac{\partial}{\partial \Vec{r}^{*}} \mathbf{K}^\text{OQML} 
        \end{bmatrix} \bm{\alpha}^\text{OQML}
    \end{equation}
    where $\mathbf{K}^\text{OQML}$ is the matrix of kernel elements between the atoms in the training set and the training or query molecules, and $-\tfrac{\partial}{\partial \Vec{r}^{*}}$ is the force operator.
    The presence of an asterisk in the operator denotes that that the differentation is wrt.~a coordinate in the training/query compound, while the absence of an asterisk denotes that the differentiation is carried out wrt.~the coordinate of an atom/molecule used to form the basis set.
    Thus, the dimension of the OQML kernel is $\left(3MN+N\right) \times MN$ where $N$ is the number of molecules in the training set, and $M$ is the average number of atoms in each molecule. 
    
    A solution to Eq.~\ref{eq:oqml} can be obtained by minimizing the following cost function:
    
\begin{equation} \label{eq:lstsq2}
J(\bm{\alpha}^\text{OQML}) = \norm{ \begin{bmatrix}
            \mathbf{U} \\
            \mathbf{F} 
        \end{bmatrix} - \begin{bmatrix}
            \mathbf{K}^\text{OQML} \\
            -\frac{\partial}{\partial \Vec{r}^{*}} \mathbf{K}^\text{OQML} 
        \end{bmatrix} \bm{\alpha}^\text{OQML}
        }^2_2
\end{equation}
    with respect to $\bm{\alpha}^\text{OQML}$.
    This least-squares approach leads to a solution that looks similar to the normal equation.
    However, we found that this approach involves the product of kernel matrices that are ill-conditioned and therefore suffers from numerical instability, leading to large training and and test errors.

    A more numerically stable approach involves solving Eq.~\ref{eq:oqml} directly, using a singular-value decomposition (SVD), which is used for OQML  in this work.
    In similar spirit to the regularization used in KRR, the smallest singular values (below a certain threshold) can be ignored in the solution.
    This threshold, $\epsilon_\mathrm{min}$, can be treated as a hyperparameter in the model.

    The elements of $\mathbf{K}^\text{OQML}$ are given by:
    \begin{equation}
           \mathbf{K}^\text{OQML}_{iJ} = \sum_{I \in i}  \mathpzc{k}\left(\mathbf{q}_J, \mathbf{q}^{*}_I\right)\label{eq:oqml_energy_kernel}
    \end{equation}
    where $J$ is an atom in the training set, and $I$ is an atom in molecule $i$.
    In contrast to the kernel matrix in KRR, $\mathbf{K}^\text{OQML}$ is non-square and has a column for each of the atoms in the training set and a row corresponding to each of the molecules in the training or query set.
    %

    The kernel matrix elements that correspond to the atomic forces are calculated by taking the negative derivative of the matrix elements in Eq.~\ref{eq:oqml_energy_kernel} with respect to the coordinates of the query molecules, that is:
    \begin{equation}
           -\frac{\partial}{\partial \Vec{r}_{K}^{*}} \mathbf{K}^\text{OQML}_{iJ} = - \sum_{I \in i} \frac{\partial \mathpzc{k}\left(\mathbf{q}_J,\mathbf{q}^{*}_I\right)}{\partial \Vec{r}_{K}^{*}}
    \end{equation}
    where $\Vec{r}^{*}_{K}$ denotes the $K$'th coordinate of the query molecule.
    The resulting derivative kernel thus has a column for each of the atoms in the training set and a row corresponding to each of the gradient components in the training or query set.
     
    %

    Energies are predicted from the set of $\bm{\alpha}$-coefficients:
    \begin{equation}
            \mathbf{U} = \mathbf{K}^\text{OQML} \bm{\alpha}^\text{OQML}
    \end{equation}
    The force prediction is simply the derivative of the above equation with the same set of $\bm{\alpha}$-coefficients:
    \begin{equation}
            \mathbf{F} = -\frac{\partial}{\partial \Vec{r}^{*}} \mathbf{K}^\text{OQML} \bm{\alpha}^\text{OQML}
    \end{equation}
    See Appendix B for the derivation of all kernel derivatives mentioned in this section.
    %
    
    In contrast to methods that learn forces directly as a vectorial quantity,\cite{CovariantKernelsSandro2016,MLatoms_2015} the use of the force operator guarantees that the machine learned potential will describe a conservative force field.
    This property is crucial for applications in molecular dynamics where energy conservation is necessary to obtain correct sampling without heavy use of thermostats.

    \subsubsection{Gaussian Process Regression Including Derivatives}
    It is also possible to define models that incorporate derivatives in the training set within the framework of Gaussian process regression.\cite{RasmussenWilliams}
    The relevant equations for training a model on energies and forces for chemical compounds are presented below.
    For their derivation we refer the reader to the work of Mathias,\cite{sonjamathias} and the work of Bart\'ok and Cs\'anyi.\cite{GAPtutorial}
    The Gaussian process regression kernel matrix which simultaneously incorporates the energy, $\mathbf{U}$, and the forces, $\mathbf{F}$, is written as:
    \begin{equation}\label{eq:gpr_derivative}
        \begin{bmatrix}
            \mathbf{U} \\
            \mathbf{F} 
        \end{bmatrix} = \begin{bmatrix}
           \mathbf{K}^\text{KRR} && -\frac{\partial}{\partial \Vec{r}}\mathbf{K}^\text{KRR} \\
           -\frac{\partial}{\partial \Vec{r}^{*}}\mathbf{K}^\text{KRR} && \frac{\partial^2}{\partial \Vec{r}\partial \Vec{r}^{*}}\mathbf{K}^\text{KRR} 
        \end{bmatrix} \bm{\alpha}^\text{GPR}
    \end{equation}
    where $\mathbf{K}^\text{KRR}$ is the same kernel matrix as used in KRR, as described previously. 
    In the following, we abbreviate the above methodology of Gaussian process regression derivatives with as "GPR".

    The first of the two off-diagonal blocks contain only one derivative given by
    \begin{equation}
           -\frac{\partial}{\partial \Vec{r_{K}^{*}}} \mathbf{K}^\text{KRR}_{ij} =
           - \sum_{I \in i}\sum_{J \in j} \frac{\partial \mathpzc{k}\left( \mathbf{q}_J, \mathbf{q}^{*}_I\right)}{\partial \Vec{r}_{K}^{*}}
    \end{equation}
    where $\Vec{r}^{*}_{K}$ denotes the $K$'th coordinate of the query compound. The other block is given analogously.
    The last block which comprises the largest part of the GPR kernel matrix is the double derivative given by:
     \begin{equation}
           \frac{\partial^2}{\partial \Vec{r_{L}}\partial \Vec{r_{K}^{*}}}\mathbf{K}^\text{KRR} _{ij} =
     \sum_{I \in i}\sum_{J \in j} \frac{\partial \mathpzc{k}\left(\mathbf{q}_J, \mathbf{q}^{*}_I\right)}{\partial \Vec{r}_{L}\partial \Vec{r}_{K}^{*}}\label{eq:second_derivative}
    \end{equation}
    where $\Vec{r}_{L}$ and $\Vec{r}^{*}_{K}$ denotes the $L$'th and $K$'th coordinate of the basis and query compounds, respectively.
    
    %
    Thus, the dimension of the GPR kernel is $\left(3MN+N\right) \times \left(3MN+N\right)$ where $N$ is the number of molecules in the training set, and $M$ is the average number of atoms in each molecule.
    The rows of the full GPR kernel matrix thus run over the same indices as the the OQML kernel matrix.
    However, where the indices of the columns of the OQML kernel matrix run over the atoms in the training set, the indices of the columns GPR kernel run first over the molecules in the training set and secondly over the gradient components in each molecule.
    Thus, the main difference is the choice of basis in which the regression problem is expanded.
    
    The regression coeffients $\bm{\alpha}^\text{GPR}$ can be obtained by minimizing a cost function similar to that in Eq.~\ref{eq:krr_lagrangian}, only with the difference that the matrix $\mathbf{K}^\text{KRR}$ is replaced by the GPR kernel matrix in  Eq.~\ref{eq:gpr_derivative}, and $\mathbf{U}$ is replaced by the vector that contains both the energies and atomic forces.

    Compared to OQML, the GPR kernel matrix contains derivative terms of up to second order, whereas OQML only contains terms up to first order.

    The second-order part of the GPR kernel matrix is the computationally heaviest term, and the time to compute it scales as  $\mathcal{O}\left(36N^2M^4\right)$, where $N$ is the number of molecules in the training set, and $M$ is average number of atoms in each molecule.\cite{christensen2018operators}
    In comparison, the heaviest term of the OQML kernel is only of first order and scales as $\mathcal{O}\left(6N^2M^3\right)$.\cite{christensen2018operators}
    Both methods scale as $\mathcal{O}\left(k N^2\right)$ with the number of training molecules, but GPR has a higher prefactor and scales much less favorably with the number of atoms in the individual molecules (quartic rather than cubic).

    We also note the existence of sparsification procedures which can be applied to the GPR model, such as those used within the Gaussian approximation potential (GAP) methods.\cite{GAPtutorial}
    The use of sparsification makes it possible to treat the problem without any second derivatives in the kernel matrix.
    The result here is then a kernel matrix that is very similar to the OQML kernel and requires far less time to compute compared to the full GPR kernel.
    
    As the molecules used to benchmark force prediction methods in this study contain between 9-21 atoms, it is expected that the time to calculate the kernel for GPR is on the order of 50-200 times slower than OQML.
    This is demonstrated numerically in the Timings section \ref{sec:timings}.

    Additionally we note that evaluation of the second derivative matrix elements in GPR scales as $\mathcal{O}\left(N^2\right)$ with regards to the size of the representation, while evaluation of first derivative matrix elements (such as in OQML and the off-diagonal blocks in GPR) only scale as $\mathcal{O}\left(N\right)$, see for example Eq.~\ref{eq:firstderiv} and \ref{eq:secondderiv} in the Appendix B.
    
    In terms of memory usage, the GPR kernel is roughly three times larger than the OQML kernel, since it contains a column for each molecule and gradient component in the training set, whereas the OQML kernel only contains a column for each atom in the training set.
    As a result, GPR scales computationally less favorably compared to OQML, although the accuracy of the regression may be slightly increased due to more regression coefficients being fitted.

    \subsubsection{Gradient-Domain Machine Learning (GDML)}
    Since we will be comparing numerical results from the GDML\cite{chmiela2017machine} and the closely related sGDML\cite{Chmiela2019sGDML} methods, we also briefly review these approaches for the sake of completeness.
    GDML can be seen as equivalent to the GPR approach detailed above, with the difference that the energy is left out of the training data, such that only forces are used in the training.
    In turn, the corresponding 0th and 1st derivative kernel blocks from the GPR kernel are not present in the GDML kernel.
    Thus the kernel in the GDML approach is identical to the block in the GPR kernel which contains the second-order derivative.
    Effectively, the equation solved in the GDML-approach is:
    \begin{equation}\label{eq:gdml_force}
            -\mathbf{F} = \frac{\partial^2}{\partial \Vec{r^{}}\partial \Vec{r}^{*}}\mathbf{K}^\text{KRR} \bm{\alpha}^\text{GDML}
    \end{equation}
    The GDML regression coefficients can be obtained, similarly to those in GPR and KRR, by minimizing a cost function similarly to that in Eq.~\ref{eq:krr_lagrangian}, but only including the forces and second-derivative kernel matrix in the above equations.

    Force predictions are then calculated using Eq.~\ref{eq:gdml_force}, while energy predictions in the GDML approach are done using a 1st derivative kernel:
    \begin{equation}
        \mathbf{U} = \frac{\partial}{\partial \Vec{r}}\mathbf{K}^\text{KRR} \bm{\alpha}^\text{GDML} 
    \end{equation}
    %
    Note that this derivative is taken with respect to the basis and not the query molecule.
    %
    %
    Since the energy is not used in the training step, all predicted energies are only defined up an integration constant which can be inferred from the mean deviation between predicted energies for a training or validation set.
    
    Compared to GPR, the computational cost of GDML is ever so slightly reduced, as the three smaller blocks are being ignored in the kernel.
    However, the corresponding gain in computational cost is negligible.
    Leaving out the energy in the training set makes it difficult to regress any energy offset if a model is trained across chemical composition and molecular size.
    For GDML models that are only trained on one molecule, however, this seems to have very little effect.\cite{chmiela2017machine}
    
    In the formulation of GDML and sGDML by Chmiela \textit{et al.},\cite{chmiela2017machine,chmiela2017machine} one further performance enhancement is made, compared to the equations mentioned for KRR and GPR herein.
    Instead of using a representation for each atomic environment, GDML and sGDML both use one "global" representation for the entire molecule.
    In GDML, the inverse interatomic distance matrix is used, while sGDML is a variant of GDML which takes atoms with symmetry into account. 
    Other notable global representations are the Coulomb matrix,\cite{RuppPRL2012} BoB,\cite{AssessmentMLJCTC2013} SLATM,\cite{Bing2016} the Fourier series of atomic radial distribution functions,\cite{FourierDescriptor} and the constant size descriptor of Collins \textit{et al.}\cite{constsize2018}

    The use of such global representations reduces the double sum over atomic contributions in Eq.~\ref{eq:second_derivative} to the evaluation of just the single "global" kernel derivative.
    %
    The result is a massive reduction in the evaluation speed on the order of $M^2$, where $M$ is the number of atoms in each molecule.

    While this speed up is desirable, we note that, based on our observation in Section \ref{section:results} and our results for force and energy predictions, such global representations generally display lower predictive accuracy, especially for condensed-phase systems.
    There is some evidence that global representations are not ideal for learning size-extensive properties, such as energies, since the resulting kernel elements do not scale with the size.\cite{amons2017}
    This is naturally taken into account by kernels that sum over atomic contributions.
    For example with a local representation, a system with two of the same molecule infinitely apart will naturally have twice the energy of a system containing the same molecule only once, whereas the same will not necessarily be true for a global kernel.

    Future development of global representations could potentially be fruitful due to their computational efficiency.

    We note that the speedup obtained from using global representations is similar to the difference in scaling cost between OQML and GPR, and the difference between our combined FCHL19/OQML model and a GDML-type model based on a global representation will likely be less than a factor $M$.
    
    %
    
    \subsubsection{Kernel Function}
    We introduce a variant of the Gaussian kernel function, augmented with an elemental screening function that only compares representations for atomic environments of atoms of the same element type:
    \begin{equation}\label{eq:kernel_screening}
    \mathpzc{k}(\mathbf{q}_I,\mathbf{q}_{J}^{*}) = \delta_{Z_I Z_{J}^{*}}  \exp\left(-\frac{\| \mathbf{q}_I - \mathbf{q}_{J}^{*} \|^2_2}{2\sigma^2}\right)
    \end{equation}
    where $\delta$ is the Kronecker delta function and the subscripts $Z_I$ and $Z_{J}^{*}$ are the nuclear charges of the atoms $I$ and $J^{*}$, respectively.
    The $\delta_{Z_I Z_{J}^{*}}$ term ensures that only relevant pairs of atoms are compared. 
    For example, it is likely to be of little relevance to compare the atomic environment of a carbon atom to that of a hydrogen atom. 
    Furthermore, the Kronecker delta function reduces the cost of a kernel evaluation since the calculation of many expensive combinations of kernels and their derivatives can be skipped.
    %
    If needed, the Kronecker delta function could still be changed to a function that incorporates learning across alchemical space to increase the  learning rate, as shown in our previous work.\cite{faber2017alchemical}
    %
    %
    
    In principle, any suitable kernel function could be used besides a Gaussian function, and the choice could be treated as a hyper-parameter of the model.
    For simplicity, however, only the Gaussian kernel is used in this work.

    \section{Results}
    \label{section:results}


    \subsection{Energy Learning}
    In this section, the FCHL19 representation is used with the "universal" set of hyper-parameters fitted to energies of non-equilibrium structures (see section \ref{sec:optparams}).
    As the geometries in the QM9 and QM7b datasets used in this section are minimized wrt.~energy, we expect a slight decrease in the predictive accuracy of FCHL19 compared to if the hyper-parameters had been optimized on similarly minimized structures.
    We compare KRR models with FCHL19 to similar KRR models with FCHL18 and a number of other models from literature.
    We note that the model and data selection methodology used to obtain the various results found in the literature might differ from the 5-fold cross-validation methodology we used in this study.
    However, we assume that such differences only give rise to negligible differences in the predictive accuracy of the models.
    
    %
    %
    
    \subsubsection{Results for QM9}
\begin{figure}
        \centering
        \includegraphics[width=\linewidth]{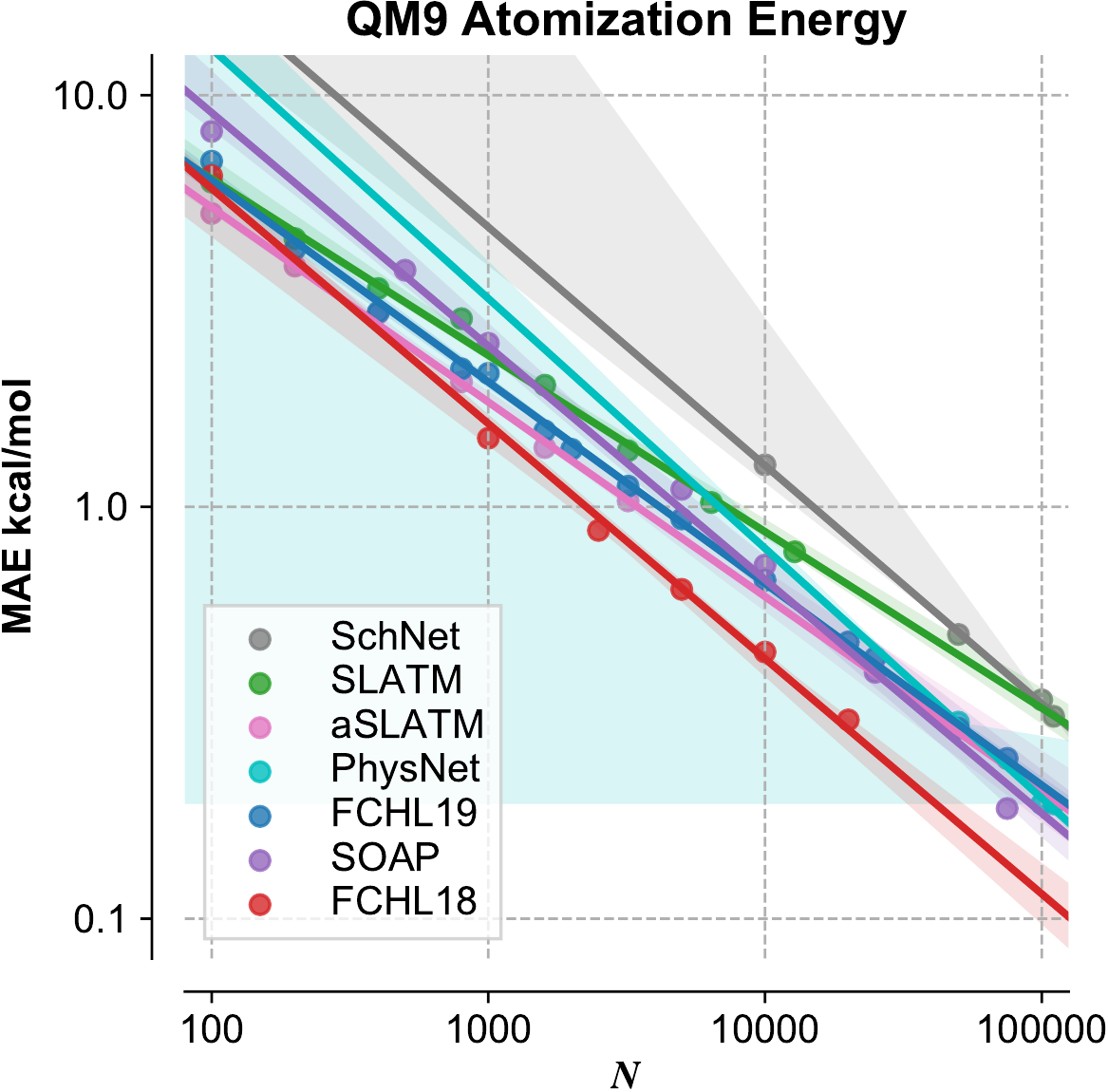}
        \caption{\label{fig:qm9} Learning curves for the QM9 dataset: The mean absolute error (MAE) of atomization enery prediction is plotted for 5 KRR models based on different representations and one neural network versus the training set size (see text). Linear fits are displayed for clarity, and shaded areas denote the 95\% confidence intervals for the fits as obtained via boot-strapping.\cite{seaborn}
            }
    \end{figure}

    In Fig.~\ref{fig:qm9} we compare the predictive accuracy of a number of kernel-based models for the atomization energy of molecules in the QM9 dataset.\cite{DataPaper2014}
    We compare FCHL19 to five other well-performing representations: the SOAP multi-kernel model,\cite{Sandip2016,BartokGabor_Descriptors2013} SchNet\cite{schutt2018schnet} and PhysNet\cite{UnkePhysNet2019} which are two the best performing neural networks for this dataset, SLATM and aSLATM,\cite{Bing2016} where the former uses one global representation for the entire molecule and the latter uses an atomic decomposition of the kernel, and finally the previous FCHL18\cite{faber2017alchemical} representation.

    For the QM9 dataset we find models based on FCHL19 to be among the models with the lowest out-of-sample MAE atomization energy predictions.
    Compared to the best performing model, FCHL18, the MAE at 20,000 training samples are 0.30 and 0.47 kcal/mol for FCHL18 and FCHL19, respectively.
    For the largest training split (75,000 training samples) the MAE for the FCHL19 model is 0.25 kcal/mol.
    %
    %
    Overall, we find that our previous FCHL18 model has the lowest prediction MAE, while SOAP, FCHL19, and aSLATM have virtually indistinguishible MAE. Finally the Global SLATM model and SchNet perform a bit worse for QM9.

    We reiterate again at this point that the hyperparameters of FCHL19, in contrast to some other models, have not been optimized on the QM9 dataset.
    
    \subsubsection{Results for QM7b}
    
    \begin{figure}
        \centering
        \includegraphics[width=\linewidth]{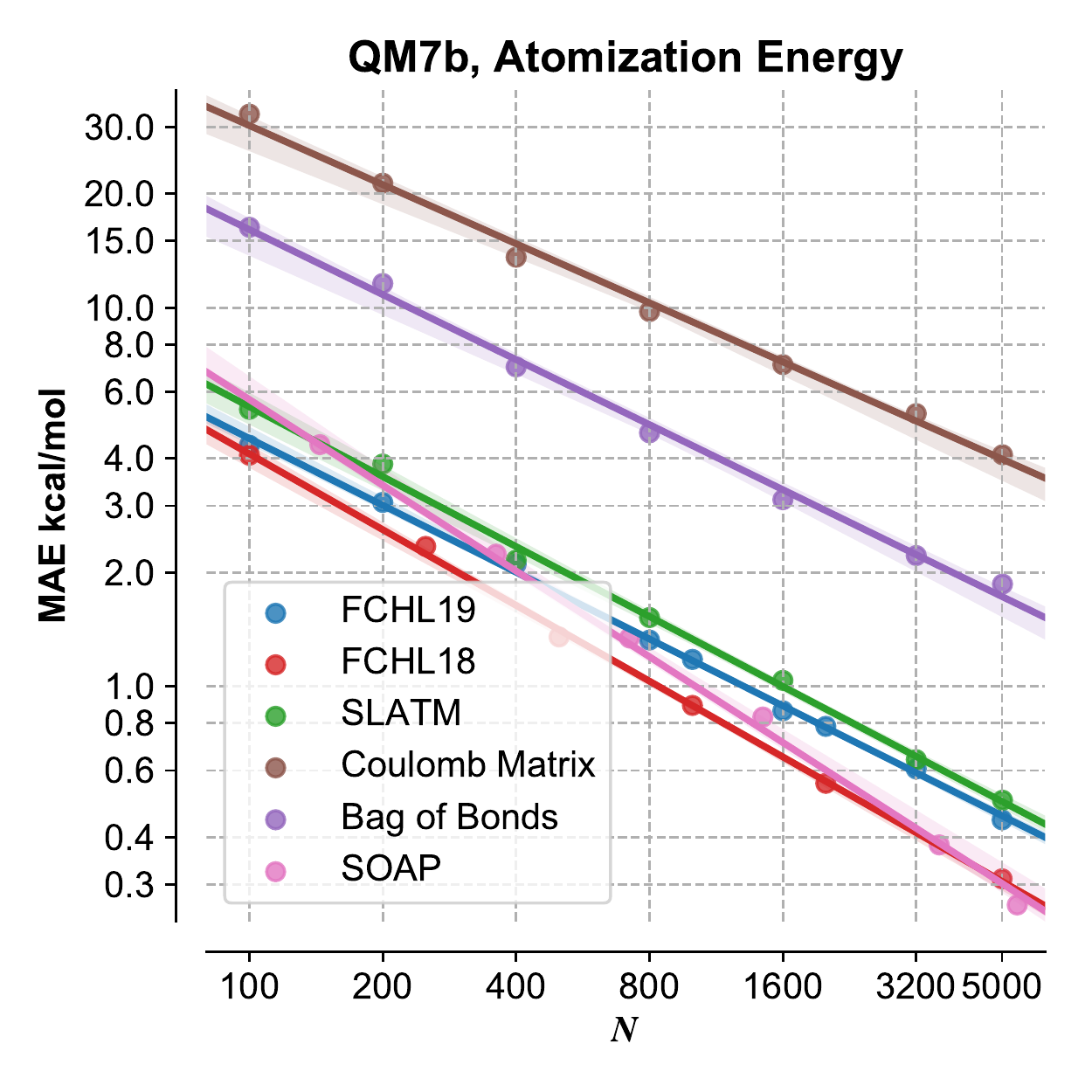}
        \caption{\label{fig:qm7b} Learning curves for QM7b: The mean absolute error (MAE) is plotted for KRR models with 6 different representations (see text) versus the training set size. Linear fits are displayed for clarity, and shaded areas denote the 95\% confidence intervals for the fits as obtained via boot-strapping.\cite{seaborn}
            }
    \end{figure}
    Similarly, Fig.~\ref{fig:qm7b} compares the predictive accuracy of a number of kernel-based models for the atomization energy of the QM7b dataset.\cite{Montavon2013}
    We compare our model to the following representations: FCHL18,\cite{faber2017alchemical} SLATM,\cite{Bing2016} the Coulomb matrix,\cite{RuppPRL2012} Bags-of-Bonds (BoB),\cite{AssessmentMLJCTC2013} and finally SOAP\cite{Sandip2016,BartokGabor_Descriptors2013}.
    Data for these models are obtained from Ref.~\citen{faber2017alchemical}.

    As expected, the dataset is too small for Coulomb matrix and BoB to reach chemical accuracy. 
    In contrast, all other models (FCHL19, FCHL18, SLATM, and SOAP) reach chemical accuracy when trained on between 800 and 1,600 samples. 
    Additionally, the fitted learning curves (Fig.~\ref{fig:qm7b}) display similar predictive accuracies.
    For example, all these models are within an MAE of $\pm$0.3 kcal/mol of FCHL19 at 1,000 training samples.
    %
    %
    \subsubsection{Results for QM7b-T and GDB13-T}
    
    \begin{figure*}
        \centering
        \includegraphics[width=0.45\linewidth]{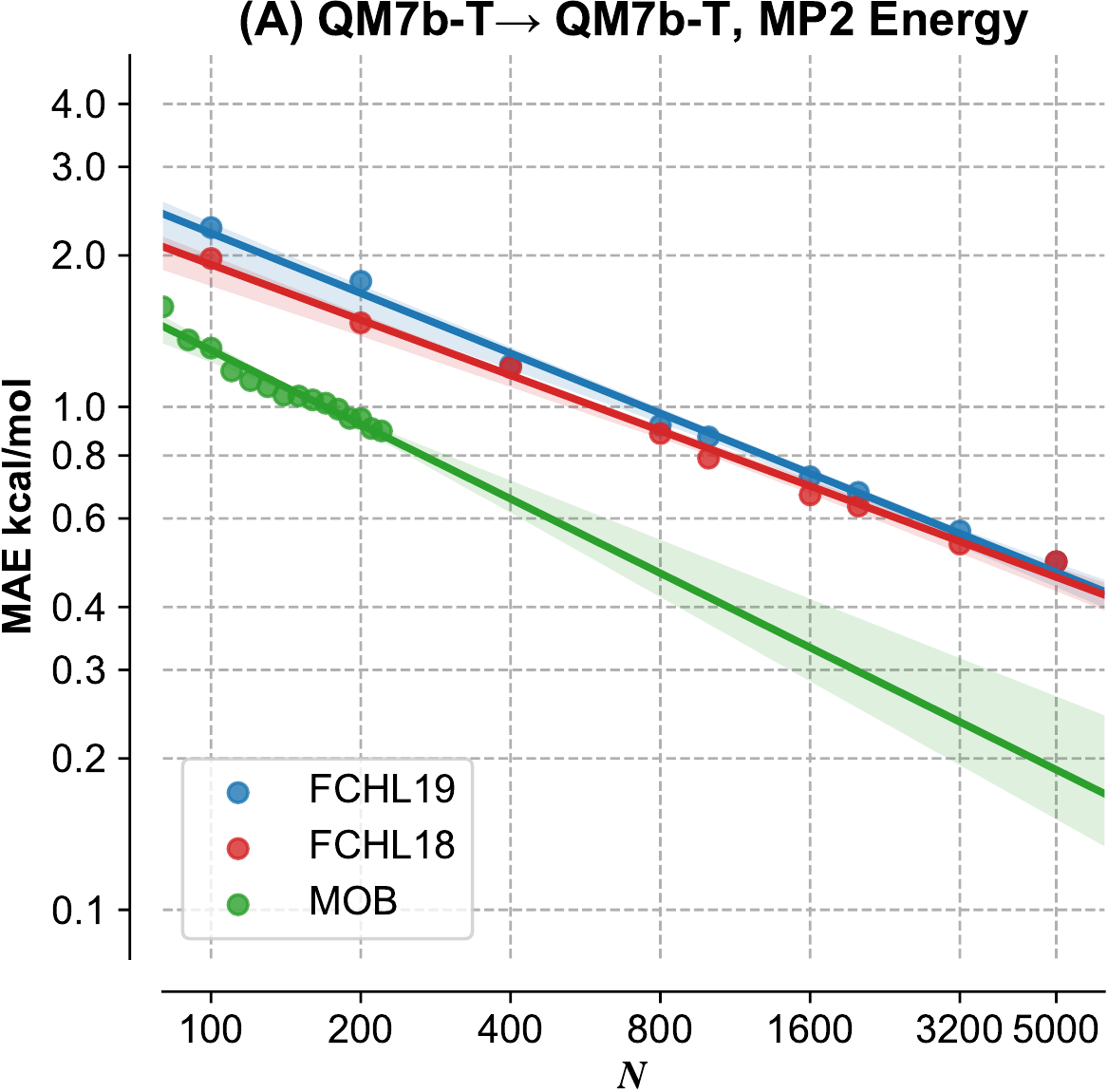}
        ~~~~~~~~~
            \includegraphics[width=0.45\linewidth]{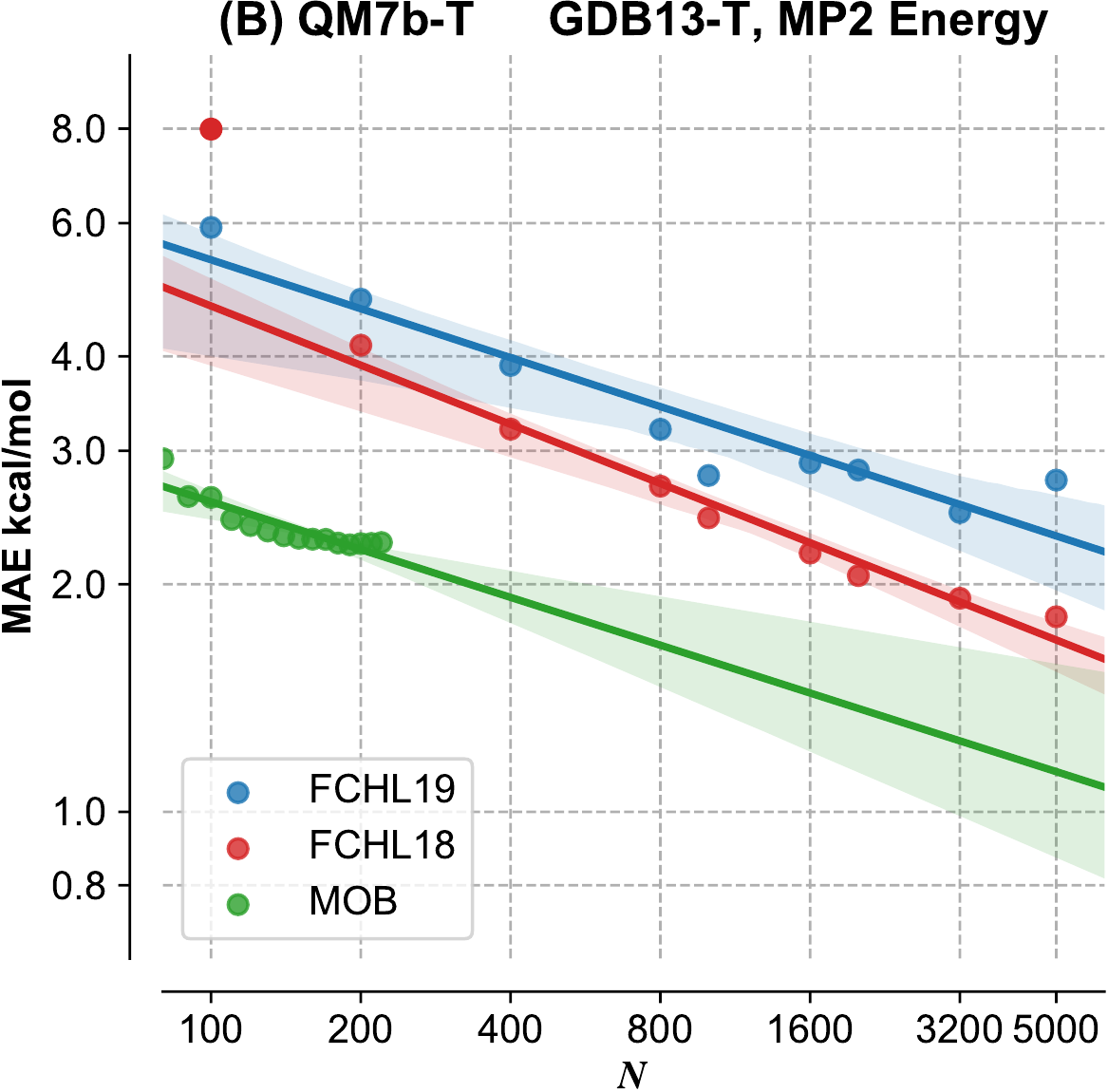}
            \caption{\label{fig:qm7bt}
        The two figures display the out-of-training error for models trained on subsets of the QM7b-T dataset.
        In (A) the models predict the MP2 correlation energy of unseen samples from the same QM7b-T dataset, while in (B) the same models predict the energy on unseen samples from a subset of GDB13-T dataset. Linear fits are displayed for clarity, and shaded areas denote the 95\% confidence intervals for the fits as obtained via boot-strapping.\cite{seaborn} For FCHL18, the fit and boot strapping is performed without including the first data point as the fit would otherwise appear unreasonably steep.
                }
    \end{figure*}
    
    While the QM7b and QM9 datasets contain energies for the equilibrium geometry of small molecules, the QM7b-T\cite{mobml2} and GDB13-T\cite{mobml2} datasets contain non-equilibrium geometries of molecules from QM7b\cite{Montavon2013} and GDB-13.\cite{ReymondChemicalUniverse3}
    In addition to gauging the accuracy of a model on unseen samples from the same dataset by training and predicting on subsets of QM7b-T, we also benchmark how well a model can extrapolate to prediction on samples from a dataset containing larger molecules by predicting on GDB13-T with models trained on QM7b-T.
    Learning curves for these tests can be seen in Fig.~\ref{fig:qm7bt}.
    
    First, we compare FCHL19 to FCHL18 and the Molecular-Orbital-Based machine learning method (MOB)\cite{mobml1,mobml2} by training on the QM7b-T dataset and predicting on unseen samples from the same dataset.
    FCHL19 and FCHL18 both reach 1 kcal/mol accuracy for this dataset at between 400 and 800 training samples, and above 1,000 the difference is less that 0.1 kcal/mol, with FCHL18 being consistently slightly more accurate.
    The MOB method, which requires a Hartree-Fock calculation for every query to calculate the localized molecular orbitals used to generate the representation, reaches 1 kcal/mol at about 200 training samples and is consistently more accurate with about a 2-3 times improvement in accuracy.

    Secondly, we test the extrapolative power of the three models by training models on the QM7b-T dataset and predicting on the GDB13-T dataset.
    In this test, the differences observed previously seem to be magnified.
    Neither the FCHL19 nor the FCHL18 models reach chemical accuracy for GDB13-T dataset with the amount of training data available in the QM7b-T dataset.
    At 1,000 training samples, the MAE for the two models are 2.7 and 2.2 kcal/mol, respectively. 
    In comparison, MOB reaches this error at around 100-200 samples, however a larger MOB training set is unavailable due to the difficulty of training large models for MOB.\cite{mobml1,mobml2}
    

    \subsubsection{Results for Water40}
    \begin{figure}
        \centering
        \includegraphics[width=\linewidth]{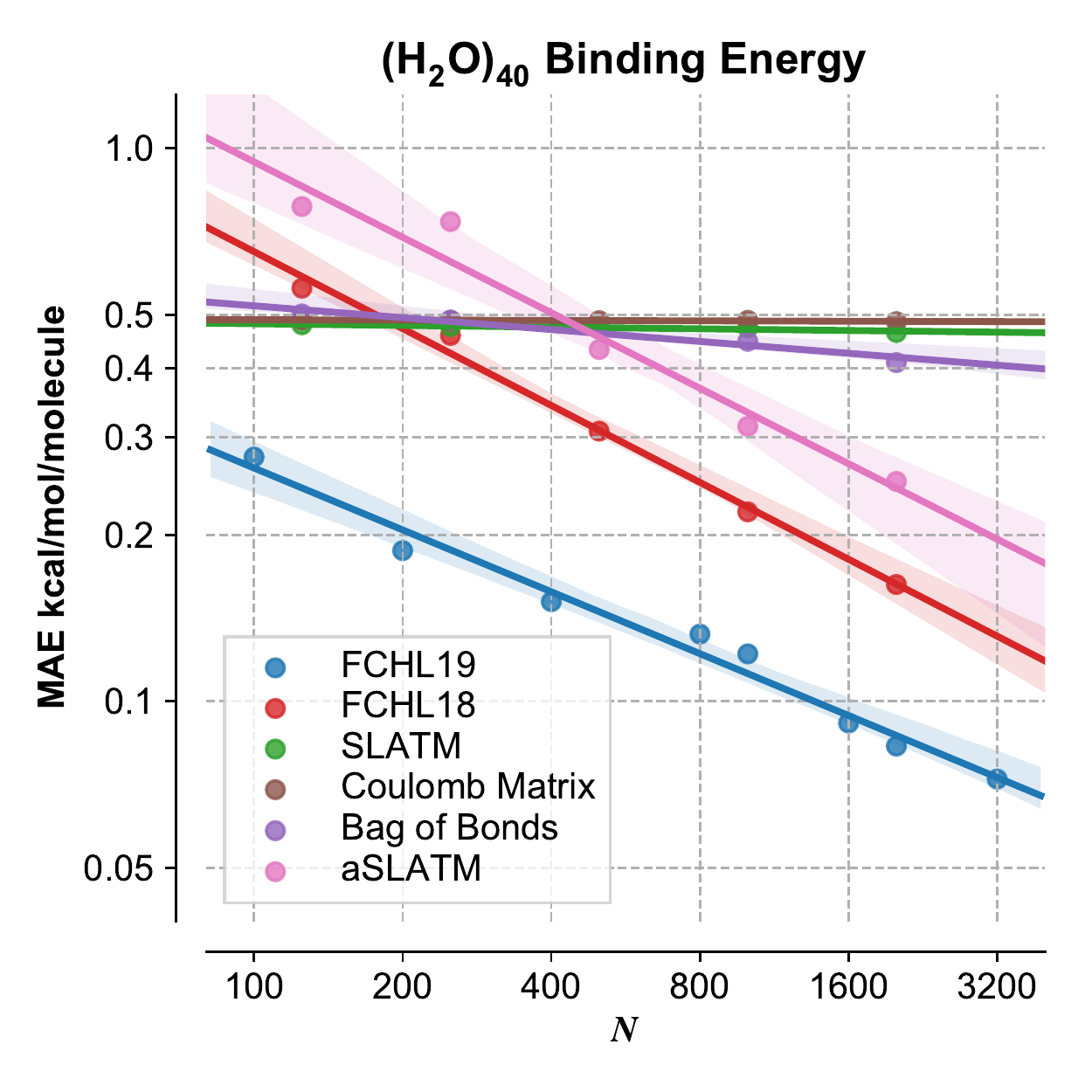}
        \caption{\label{fig:water40} Learning curves for the Water40 dataset: The mean absolute error (MAE) binding energy per molecule is plotted for 6 different representations versus the training set size. Linear fits are displayed for clarity, and shaded areas denote the 95\% confidence intervals for the fits as obtained via boot-strapping.\cite{seaborn}
            }
    \end{figure}
    The Water40 dataset consists of 10,000 MD snapshots of a water cluster with 40 water molecules for which a DFT single-point energy has been calculated.\cite{christensen2018operators}
    As such, this dataset probes the performance of ML models on chemical systems that approach the condensed phase behavior.
    Here we compare our model to the following representations: FCHL18,\cite{faber2017alchemical} SLATM and aSLATM,\cite{Bing2016} the Coulomb matrix,\cite{RuppPRL2012} and  BoB.\cite{AssessmentMLJCTC2013}
    The learning curves for these models on the Water40 dataset are displayed in Fig.~\ref{fig:water40}. 
    
    We find that the accuracy of machine learning models based on the FCHL19 representation have far greater predictive accuracy compared to any other representation. 
    For example, for models trained on 1,000 training instances, the FCHL18-based model yields an MAE of 0.22 kcal/mol/molecule, while the model trained using the FCHL19 representation yields an MAE test error of 0.12 kcal/mol/molecule.
    The FCHL19 representation reduces the data required to reach a given accuracy by roughly 5 times compared to FCHL18, and by roughly 10 times compared to aSLATM.

    Note that for Water40, and in contrast to molecular datasets, the use of global representations (i.e.~those that do not use a decomposition of the kernel in local, atomic contributions), such as the Coulomb matrix, BoB, and SLATM, results in models which hardly display any learning at all, with a constant error of about 0.5 kcal/mol/molecule, regardless of training set size.

    Although FCHL19  is parametrized for the atomization energy of small molecules, it nevertheless yields superior accuracy for the binding energy of water clusters where accurate handling of non-covalent interactions are key to determining the energy.
    This suggests that the  parameters in the representation have a high degree of transferability and do not necessarily need to be re-parametrized for every new dataset.

    While the accuracy of models based on FCHL19 is better than that of models based on other representations, we expect that models based on, for example, FCHL18 and aSLATM are likely to reach a similar accuracy if the model parameters of those representations are obtained similarly to those of FCHL19.

    \subsection{Force learning}
    In the following section the FCHL19 representation is used with parameters that are optimized for both force and energy prediction simultaneously (see section \ref{sec:optparams}).
    %
    
    \subsubsection{Results for MD17}
    \begin{figure*}
        \centering
        \includegraphics[width=\linewidth]{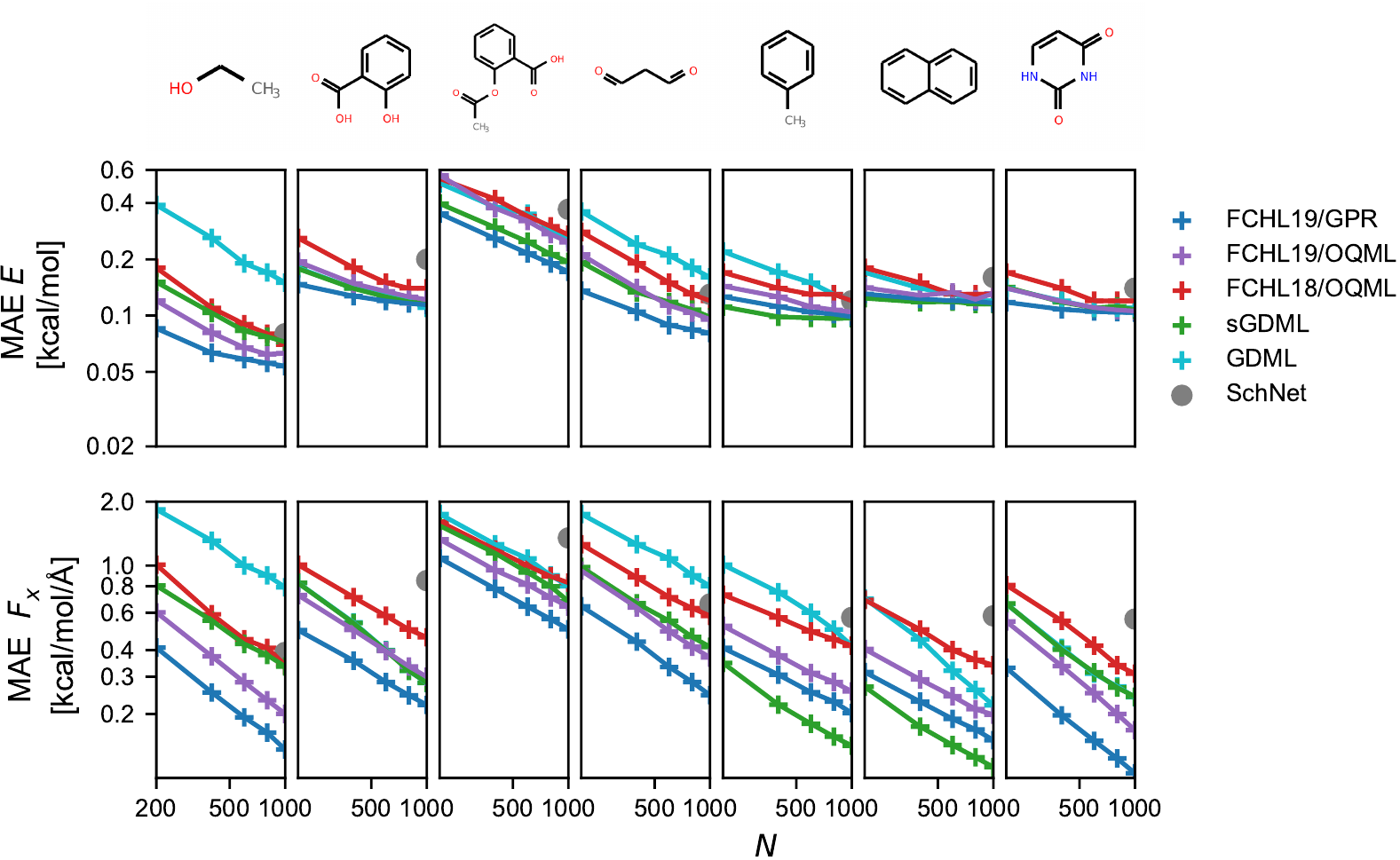}
        \caption{Here we present learning curves for force and energy learning for seven molecules from the MD17 dataset. Learning curves are presented for 6 different QML models (see text). The top row contains learning curves for the out-of-sample MAE energy prediction (MAE $E$), and the bottom row contains corresponding learning curves for out-of-sample MAE force component prediction (MAE $F_x$), for the molecules (from left to right) ethanol, salicylic acid, aspirin, malonaldehyde, toluene, naphthalene, and uracil. \label{fig:md17} 
            }
    \end{figure*}
    Fig.~\ref{fig:md17} reports the MAE force and energy prediction as a function of the number of training samples taken from 7 molecules from the MD17 dataset.\cite{chmiela2017machine}
    We note that the original MD17 dataset also includes a dataset for benzene.
    However, due to low accuracy in the reported energies for this dataset, we have chosen to exclude this from our tests, as the high noise level would be the dominating error, and as such would not reflect differences in the machine learning procedures.
    We compare OQML and GPR models based on FCHL19 to OQML models based on FCHL18.\cite{faber2017alchemical,christensen2018operators}
    %
    In addition, we compare to GDML\cite{chmiela2017machine} and sGDML\cite{Chmiela2019sGDML} which are two state-of-the-art kernel-based methods closely related to GPR.
    Furthermore, we compare to one of the best performing neural networks for forces, SchNet,\cite{schutt2018schnet} which is based on a continuous-filter convolutional neural network.

    In general we note that the reparametrized FCHL19 representation leads to models that have improved accuracy compared to the FCHL18 prediction errors reported in our previous paper.\cite{christensen2018operators}
    Learning curves for these models are presented in Fig.~\ref{fig:md17}.
    
    
    For all molecules in the MD17 dataset, the FCHL19 representation with both the GPR and OQML regressors display faster learning compared to FCHL18 with the OQML regressor for both energy and force learning.
    As a general trend, FCHL19/OQML requires about half the samples to reach a given accuracy compared to FCHL18/OQML.
    Changing to the GPR regressor, FCHL19/GPR in turn requires about half the samples to reach the same accuracy as FCHL19/OQML.
    For example, for ethanol, an MAE force error of 0.4 kcal/mol/\AA~error is obtained at roughly 200, 400 and 800 samples for FCHL19/GPR, FCHL19/OQML and FCHL18/OQML, respectively.

    Similar trends are observed for both force and energy learning for salicylic acid, aspirin, malonaldehyde, and uracil.
    For these molecules we find that FCHL19/GPR has the highest accuracy in all cases, with the sGDML method and FCHL19 with the OQML regressor also performing very well and at much reduced computational costs.
    %
    %
    
    We note that GDML and FCHL18/OQML overall have the lowest accuracy of the kernel methods, and SchNet slightly worse on average, although the time-to-train for SchNet reportedly is much more favorable for larger training sizes.\cite{schutt2019schnetpack}
    For toluene, naphthalene and uracil we note very slow energy learning for all the presented methods, with almost flat learning curves at around 0.1 kcal/mol error.
    However, this seems to be an inherent property of the dataset, and likely to be related to noise in the calculated DFT energies, for example from use of unconverged integration grids.\cite{Lejaeghereaad3000,bootsma_wheeler_2019}
    Furthermore, for the molecules toluene and naphthalene, we observe that sGDML performs very well for force learning, compared to the FCHL19 variants.
    We speculate that the comparably poor performance of variants of FCHL is due to the high degree of symmetry in the 6-membered rings of the molecules; 
    when the molecule has many atoms of the same element type at very close radial distances, the Fourier transform of the angular histogram in the three-body term become very crowded, and this might lead to slower learning for molecules containing such moieties of high symmetry.
    This might be improved upon by reoptimizing the hyperparameters specifically for this system.
    Nevertheless, FCHL19 with both the OQML and GPR regressors are within 0.1 kcal/mol energy error and and 0.1 kcal/mol/\AA~force component error of the best performing method (sGDML) at 1,000 training samples in both of these cases.
    %
    %
    
    
    \subsection{Timings}\label{sec:timings}
    
    In this section, we report timings for generating the training kernel which is the most costly step for these kernel models.
    For force predictions we additionally report the prediction time per atom of FCHL19-based models trained with the OQML regressor.
    %
    All timings in this section were carried out on a 24-core compute node equipped with two Intel Xeon E5-2680v3 @ 2.50GHz CPUs and 128GB RAM.
    
    \subsubsection{Timings for Energy Learning}
    %
    %
    Using the implementations in the \texttt{QML} software package,\cite{qmlcode2017} we compare timings for calculating kernels for three representations that all use a decomposition of the kernel into atomic contributions, namely FCHL19, FCHL18, and aSLATM.
    %
    %
    For FCHL18 and aSLATM, all parameters are set to the default values in \texttt{QML}, and for FCHL19, the values in Appendix A are used.
    These timings are given in Table~\ref{tab:timings_energy}.
    In all cases, the training times scale as $\mathcal{O}\left(N^2\right)$ with the training set size, while the prediction time scales as $\mathcal{O}\left(N\right)$.

    To illustrate the effects of elemental complexity, we compare timings for both QM7 and QM7b.
    The two datasets contain molecules with up to 7 non-hydrogen atoms, with the largest molecule being 23 atoms total in both sets, and both datasets contain about 7K molecules.
    They differ, however, in the elements that are present in the two datasets: QM7 contains HCNOS while QM7b additionally contains Cl.
    As the size of the three-body terms in aSLATM and FCHL19 representations scale cubically and quadratically, respectively, with the number of elements in the dataset, the result will be a substantial increase in kernel evaluation time for models based on these representations.
    
    For aSLATM the two datasets take 4,955s and 7,727s to compute, respectively, whereas for FCHL19 the same numbers are 216s and 310s.
    In contrast, FCHL18 is largely unaffected by chemical complexity, with kernel evaluation times of 3,164s and 3,286s for the two sets, respectively.

    Additionally, we present timings for the QM9 dataset. These timings are also presented in Table~\ref{tab:timings_energy}.
    This dataset, contains 133,855 molecules with the elements HCNOF, and molecules with up to 9 non-hydrogen atoms, where the largest molecule contains 29 atoms.
    Using the previous implementations of aSLATM and FCHL18, calculating the kernel matrix for this dataset can only be done on a reasonable timescale on cluster with several nodes.
    For aSLATM and FCHL18 the time to calculate the QM9 kernel is 728 hours and 548 hours on our 24-core node, respectively.
    In contrast, for FCHL19, the time to calculate the kernel is 27 hours on the same node.
    The speedup compared to aSLATM comes from the reduced size of the representation and the element-wise kernel function which is not normally used with aSLATM.\cite{Bing2016}

    Note, that these timings only cover calculating the training kernel, and not the representation generation or regression solver.
    Generating the representations scales as $\mathcal{O}\left(N\right)$ with the number of training or predictions samples and is insignificant in comparison.
    While solvers to obtain the regression coefficients typically scale as $\mathcal{O}\left(N^3\right)$ with the number of training samples, this step is in practice insignificant compared to generating the kernel, even for the largest kernels due to a lower prefactor.
    For example, the \texttt{QML} software package uses a Cholesky decomposition as implemented in libraries such as Intel Math Kernel Library (MKL), and using this implementation for the largest kernel studied in this section (QM9) this step takes less than one hour, whereas the time to generate the kernel takes between 27 to 728 hours.

    \subsubsection{Timings for Force Learning}\label{sec:timings_force}
\begin{figure}
        \centering
        \includegraphics[width=\linewidth]{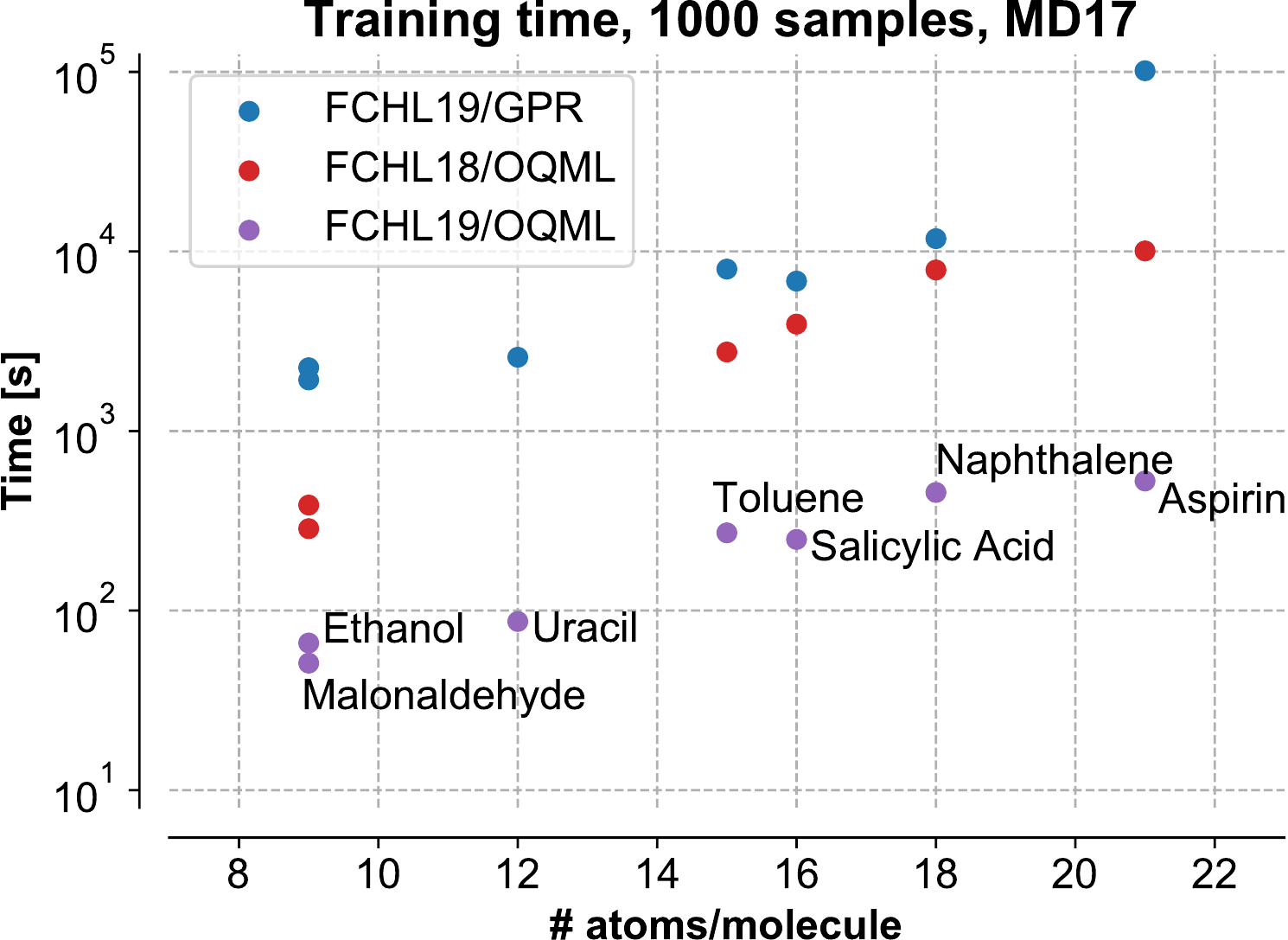}
        \caption{\label{fig:timing}The time to calculate the training kernel for 1,000 training samples with three different methods is displayed for 7 molecules from the MD17 dataset, namely ethanol, malonaldehyde, uracil, toluene, salicylic acid, naphthalene, and aspirin.
        Timings are displayed for the methods FCHL19/GPR, FCHL18/OQML, and FCHL19/OQML, and are calculated as averages over 5 kernels using different random splits, run on a 24-core node equipped with two Intel Xeon E5-2680v3 @ 2.50GHz CPUs. 
            }
    \end{figure}

    Next, we report timings for kernel evaluations for calculating the training kernel for force and energies for a set of 1K molecules taken from the MD17 dataset.
    These timings are given in Table \ref{tab:timings_forces}.
    Again, in all cases, the training times scale as $\mathcal{O}\left(N^2\right)$ with the training set size, while the prediction time scales as $\mathcal{O}\left(N\right)$.
    Compared to the FCHL18 representation, the speedup using the same regressor (OQML) is as low as 5 times for the smallest molecules, ethanol and malonaldehyde, and up to almost 20 times for the largest molecule, aspirin. 
    %
    %
    For models based on FCHL19 with the OQML regressor, the training times vary between around 51 seconds for malonaldehyde and 527 seconds for aspirin.
    These numbers correspond to force prediction times (also given in Table \ref{tab:timings_forces}, with a graphical overview in Fig.~\ref{fig:timing}) in the range of 5.7 to 25.3 milliseconds per atom for models trained on 1,000 training samples, excluding generation of the representation.
    The time to generate the representations and Cartesian derivatives of the representation was found to be very negligible in comparison: for one of the two smallest molecules in MD17, namely ethanol with 9 atoms, the time was found to be 0.27ms per atom, while for the largest molecule (aspirin with 21 atoms), the time to compute the representation was found to be 1.0ms per atom.
    
    Models based on FCHL19 with the GPR regressor are found to be substantially slower than OQML models.
    For the smallest molecules (ethanol, malonaldehyde, and uracil), the GPR kernel can be calculated in less than one hour (between 1,926 s to 2,576 s), about 30-38 times slower than the corresponding OQML kernel.
    For the largest molecule, aspirin, the differences are even larger: the GPR kernel takes 101,451 seconds, 192 times slower than the corresponding OQML kernel.
    Based on the observations in this section, a GPR model requires about half the amount of training data to reach the same accuracy as a model based on OQML.
    With the $\mathcal{O}\left(N^2\right)$ scaling of both GPR and OQML, this translates to a 4 times increase in prediction speed, and consequently OQML models will be about 10-50 times faster than a GPR model if the models are trained to the same accuracy.
    This underlines how OQML is a favorable alternative to GPR, although the learning curve offsets are somewhat larger.

    \begin{table}
          \centering
         \caption{Timings for kernel evaluation for the QM7b and QM9 datasets, with the three different atomic representations, aSLATM, FCHL18, and FCHL19.
         To illustrate the effects of molecular size and elemental complexity on the kernel evaluation time, data for the three datasets QM7, QM7b, and QM9 are presented. 
         Timings are presented in seconds (s) or hours (h). Additionally, the size of each dataset and the elements present in the datasets are listed.
         All calculations are done on a 24-core node equipped with two Intel Xeon E5-2680v3 @ 2.50GHz CPUs.\label{tab:timings_energy}}
      \begin{tabular}{lrcrrr}
        \hline
        Dataset & Molecules & Elements & aSLATM & FCHL18 & FCHL19\\
        \hline
        QM7   & 7 165 & H C N O S    &  4 966 s              & 3 164 s         &  216 s \\
        QM7b   & 7 211 & H C N O S Cl     & 7 727 s               &     3 286 s       & 310 s\\
        QM9  & 133 885 & H C N O F    & 728 h          &   548 h         &  27 h\\
        \hline
      \end{tabular}
    \end{table}

    \begin{table}
         \centering
         \caption{Training times for calculating the kernel matrix for 1,000 molecules (forces and energies) of 7 molecules from the MD17 dataset with the FCHL18 and FCHL19 with the GPR and OQML are given in seconds.
         Additionally, the time to calculate the prediction kernel for FCHL19/OQML is given in ms per atom. 
         The numbers in this table are calculated as averages over 5 kernels using different random splits, run on a 24-core node equipped with two Intel Xeon E5-2680v3 @ 2.50GHz CPUs. 
         \label{tab:timings_forces}}
      \begin{tabular}{lcrrrr}
        \hline
         Molecule  &Atoms & FCHL18 & FCHL19      & FCHL19  & FCHL19\\
        & & OQML & GPR      & OQML  & OQML\\
        & & [s] & [s]      & [s]  & [ms/atom]\\
        \hline
        Ethanol        &  9 &     387 &   2 252    &  66  &   7.3 \\
        Malonaldehyde  &  9 &     286 &   1 926    &  51  &   5.7 \\
        Naphthalene    & 18 &   7 886 &  11 782    & 455  &  25.3 \\
        Aspirin        & 21 &  10 067 & 101 451    & 527  &  25.1 \\
        Salicylic Acid & 16 &   3 940 &   6 836    & 249  &  15.6 \\
        Toluene        & 15 &   2 755 &   7 976    & 271  &  18.1 \\
        Uracil         & 12 &     N/A &   2 576    &  87  &   7.3 \\
        \hline
      \end{tabular}
    \end{table}
 
    \section{Methodology}
    \subsection{Datasets}
    This section contains a brief description of the datasets used to benchmark QML models trained with the revised FCHL19 representation.
    
    \subsubsection{QM7b}
    The QM7b dataset\cite{Montavon2013} is based on a subset of the GDB-13 database,\cite{ReymondChemicalUniverse3} and consists of 7,211 molecules with up to 7 atoms of the elements CNOSCl, saturated with hydrogen atoms.
    For each molecule, the Perdew–Burke-Ernzerhof (PBE) equilibrium geometry is available along with 13 different properties also calculated at the DFT level.
    
    \subsubsection{QM9}
    The QM9\cite{DataPaper2014} dataset is similar to QM7b, only it is based on a subset of GDB-17 database.\cite{GDB17}
    In contrast to QM7b, the QM9 dataset is much larger, and contains 133 885 molecules with up to 9 atoms of the type CNOF saturated with hydrogen atoms.
    For each , the B3LYP equlibrium geometry is available, and the atomization energy is used to generate the learning curves in this study.
    Similar to previous studies we leave out the "uncharacterized" subset of 3054 molecules that did not pass a geometry consistency check when then dataset was created.\cite{Ramakrishnan2014uncharacterized}

    \subsubsection{QM7b-T and GDB13-T}
    The QM7b-T and GDB13-T datasets\cite{mobml2} consist of non-equilibrium geometries sampled from \textit{ab initio} molecular dynamics simulations at 350K.
    QM7b-T contains non-equlibrium structures of molecules from QM7b, while GDB13-T contains non-equilibrium structures of a subset of the GDB-13 database\cite{ReymondChemicalUniverse3} where each molecule contains 13 atoms of the type CNOSCl and saturated with hydrogen.
    For each molecule in the two sets the MP2 correlation energy is given, i.e.~the difference between the MP2 and the HF energy.
    This set is used to test the extrapolative powers of QML models by training on the QM7b-T dataset and predicting on the GDB13-T dataset which contains larger molecules.

    \subsubsection{Water40}
    The Water40 dataset\cite{faber2017alchemical} consists of 10 000 MD snapshots from a molecular dynamics simulation of a water cluster with 40 water molecules sampled at 300K.
    For each sample, a dispersion-corrected DFT singlepoint energy is calculated at the PBEh-3c level of theory.\cite{pbeh3c}
    In contrast to other datasets used in this study, reliable treatment of non-bonded interactions in the machine learning model is required to learn these energies accurately.
    
    \subsubsection{MD17}
    The MD17 dataset\cite{chmiela2017machine} contains snapshots from \textit{ab inito} molecular dynamics on a number of small organic molecules for which reference force and energies are calculated at the DFT level.
    Out of the dataset we benchmark our models on force and energy data from the molecules ethanol, salicylic acid, aspirin, malonaldehyde, toluene, naphthalene, and  uracil.

    \subsection{Optimization of Representation Parameters}\label{sec:optparams}
    The optimal values of the parameters used to generate the FCHL19 representation for a given atomic environment are in principle hyperparameters of the model and must be re-fitted to each individual dataset to ensure optimal learning.
    However, in our experience the variances in these parameters are relatively small and show substantial transferability from dataset to dataset.
    Since the number of parameters is relatively big (nine parameters in total), the amount of work required to ensure optimal learning for a specific dataset can be substantial.

    Instead, we propose the use of two sets of "universal" default parameters that are fitted \textit{a priori}.
    To fit these, we employed a random subset of 576 distorted geometries of small molecules with up to 5 atoms of the type CNO, saturated with hydrogen atoms, for which the forces and energies have been obtained from DFT calculations.\cite{christensen2018operators}
    This dataset is publicly available (see Ref.~\citen{training_tar_bz}).

    The set was randomly divided into a training set (384 geometries) and a test set (192 geometries).
    A model was fitted on the training set, and predictions on the test set were used to minimize the following cost function with respect to the model parameters:
    \begin{equation}
       \mathcal{L}  = 0.01 \sum_i \left( U_i - \hat{U}_i \right)^2 + \sum_i \frac{1}{n_i} \| \mathbf{F}_i - \mathbf{\hat{F}}_i \|^2\label{eq:cost_function}
    \end{equation}
    where $U_i$ is the energy of molecule $i$ in the test set and $\mathbf{F_i}$ and $n_i$ are the forces and number of atoms of the same molecule, respectively.
    The minimization was performed via Monte Carlo greedy optimization, where real-type parameters are optimized by multiplying by a factor randomly chosen from a normal distribution centered on 1 with the variance 0.05, and integer-type parameters are optimized by randomly adding +1 or -1.

    Note that in order to reduce the number of free parameters, the hyperparameters are not fitted as element-specific parameters, but rather the same value is of a hyperparameter is used to generate the representation for an atomic environment, regardless of the element-type.
    The width of the Gaussian angular function, $\zeta$, was fixed to $\pi$, as this has shown to reduce the error from the Fourier expansion to be negligible.
    The distance cut-off for these "default" values was conservatively set to 8\AA.
    
    In the end, we fit two different sets of model parameters; one for energies+forces, and one for energies.
    For the latter parameters set, the term in Eq.~\ref{eq:cost_function} that includes forces was set to zero.
    The optimal values of all parameters can be found in Appendix A.
    
    \subsection{Hyperparameter Selection}
    For all learning results in section \ref{section:results}, the hyperparameters of the model (not including in the representation) were optimized using nested 5-fold cross validation (CV).
    First, the dataset was randomized and split into 5 "outer" folds using the KFold class implemented in scikit-Learn.\cite{scikit-learn}
    Secondly, for each of the five folds, the training set was again randomized and split into 4 "inner" folds. 
    Cross validation was performed on the inner folds to select optimal values for the kernel width and regularization.
    To select optimal kernel width and regularizer, a grid search was performed for $\sigma \in \{ 1, 2, 4, 8, 16, 32\}$ and $\lambda \in \{10^{-10}, 10^{-9}, 10^{-8}, 10^{-7}, 10^{-6}\}$.
    For OQML runs, instead of screening the parameter $\lambda$, the value of lowest accepted singular value (in terms of the largest singular value) was screened in the range $\varepsilon_\mathrm{min} \in \{ 0, 10^{-12}, 10^{-11}, 10^{-10}, 10^{-9}, 10^{-8}, 10^{-7}, 10^{-6}\}$.
    For datasets with energy labels, the set of $\{\lambda/\varepsilon_\mathrm{min}, \sigma\}$ with the lowest average MAE energy within the inner CV folds was selected to predict energies on the test set from the outer CV folds.
    Similarly for datasets with both force and energy labels, the set with the lowest average $\mathcal{L}$ (see Eq.~\ref{eq:cost_function}) within the inner CV folds was selected.
    \subsection{Learning Curves}
 
    Learning curves for models based on FCHL19 are presented as the average out-of-sample mean absolute error (MAE) over the five outer CV folds of the datasets.
    The leading term in this out-of-sample error is predicted to decay as $\frac{a}{N^b}$. 
    To illustrate this effect, all learning curves are displayed on a log-log scale where this decay becomes linear, and all plotted learning curves thus contain a linear fit and the 95\% confidence interval for the fit. \cite{vapnik1994learningcurves,StatError_Muller1996,QMLessayAnatole}
    The 95\% confidence interval is obtained using bootstrapping as implemented in the Python library Seaborn\cite{seaborn} which is also used to generate the plots.
    
    \subsection{Timings}
    All timings were performed on a compute node equipped with two Intel Xeon E5-2680v3 @ 2.50GHz CPUs (24 CPU cores in total) and 128GB RAM.
    The OMP parallel kernel routines in the \texttt{QML} code were compiled with the GNU Fortran compiler version 4.8 and linked to Intel MKL.
    \texttt{QML} was installed using only the default settings, similar to those of a user installing \texttt{QML} directly from the Python Package Index (PyPI).

    \subsection{Software and Software Availability}
    All machine learning calculations were performed using the open source quantum machine learning package \texttt{QML}\cite{qmlcode2017}.
    The code to reproduce the FCHL19 representation and several of the other models used in this paper as well as the relevant kernel and kernel derivative matrices can be found in the GitHub repository for \texttt{QML} at \url{https://github.com/qmlcode/qml}.

    \section{Conclusion and Outlook}
    We have presented a revised representation for chemical compounds which enables machine learning models that have state-of-the-art accuracy and much reduced computational cost in order to easily run on hardware that is accessible to most chemists.
    The representation is built on a discetization of the previously published FCHL18 model.\cite{faber2017alchemical}
    Two sets of universal parameters for the representations were fitted to an initial training set, and demonstrated to have a high degree of transferability.

    Machine learning models trained with the revised FCHL19 representation show state-of-the-art prediction accuracy on several datasets.
    For models trained on atomization energies, such as QM7b and QM9, the accuracy is better than 0.5 and 0.25 kcal/mol at the largest training sizes, while the training times are reduced by 10 to 20 times compared to FCHL18.
    For QM7 it is possible to train a model on 7K molecules in little over three minutes, while for the full set of 133,885 molecules in QM9, a model can be trained in roughly one day on a single node, compared to three weeks with our previous models.
    In general we note that some other kernel-based models also perform very well on these datasets, namely the SLATM- and SOAP-based models.

    For the Water40 data, the revised FCHL19 model reduces the predicted binding energy error to below 0.1 kcal/mol/molecule, even with the representation being optimized solely for small molecules, demonstrating the transferability of the model.

    %
    %
    
    Models trained on the MD17 dataset with the revised FCHL19 representation and the OQML or GPR regressors were found to yield models that reach state-of-the-art accuracy in force prediction while requiring 2-4 times less data compared to FCHL18 with the OQML regressor.
    %
    The computational cost of these force predictions was found to be on the scale of milliseconds per atom.
    For energy prediction on the MD17 dataset, the predictive accuracy seems to be limited by noise in the dataset, but models based on FCHL19 were found to have low energy prediction errors nevertheless.

    Our efforts are a substantial step towards both practical and transferable models that will allow the chemist to routinely train models and run molecular dynamics simulations with machine learned potentials throughout chemical space.
    These developments should be valuable for computational materials and molecular design campaigns, as well as for more interactive and immersive virtual reality simulation environments, which have recently been extended to enable users to manipulate real-time simulations of drug-ligand binding,\cite{Deeks2019} small molecule quantum chemistry,\cite{Amabilino2019,OConnor2019} and next generation digital education.\cite{Bennie2019,Ferrell2019}
    Future work will also deal with condensed-phase systems.

    \begin{acknowledgements}
%
The National Centre of Competence in Research (NCCR) Materials Revolution: Computational Design and Discovery of Novel Materials (MARVEL) of the Swiss National Science Foundation (SNSF) is acknowledged.
This material is based upon work supported by the Air Force Office of Scientific Research, Air Force Material Command, USAF under Grant No. FA9550-15-1-0026.
LAB thanks the Alan Turing Institute under the EPSRC grant EP/N510129/1 as well as support of this work through EPSRC grant EP/P021123/1.
%
%
The authors thank David R. Glowacki and the Intangible Realities Laboratory, whose open-source VR-enabled real-time simulation framework Narupa helped inspire us to investigate faster machine learning algorithms.
We further acknowledge the use of the following software: NumPy and the F2PY tool\cite{numpy}, and OpenMP \cite{openmp08}.
\end{acknowledgements}
    
    \section*{Appendix A: Optimized Representation Parameters}
    The optimal representation parameters obtained through the Monte Carlo optimization are presented in Table \ref{tab:parameters}.
        \begin{table}
         \centering
         \caption{Optimized representation parameters for FCHL19 for energy (E), and energy and forces (E+F). $n_{Rs_2}$ and $n_{Rs_3}$ are the number of bins for a pair or triplet of element types in the two- and three-body spectra, respectively. $w$ and $\eta_3$ determine the width of the radial two- and three-body distribution functions, respectively. $N_2$ and $N_3$ determine the decay of the two- and three-body scaling functions, respectively. $c_3$ is a weight factor that determines the weight of the three-body part relative to the two-body part. $\zeta$ is the width of the Gaussian functions used in the Fourier series and fixed to $\pi$. $r_\mathrm{cut}$ is the distance cut-off, here fixed to 8.0\AA.\label{tab:parameters}}
      \begin{tabular}{lcc}
        \hline
        Parameter       &       E       &   E+F \\
        \hline
        $n_{Rs_2}$                  &  22       &  24 \\
        $n_{Rs_3}$                  &  17       &  20 \\
        $w$     [\AA$^{2}$]         &   0.41    &   0.32\\
        $\eta_3$ [\AA$^{-2}$]       &   0.97    &   2.7\\
        $N_2$                       &   2.4     &   1.8\\
        $N_3$                       &   2.4     &   0.57\\
        $c_3$ [\AA$^{N_3}$]         &  45.8     &  13.4\\
        $\zeta$                     &  $\pi$    &  $\pi$ \\
        $r_\mathrm{cut}$ [\AA]      &   8.0     &   8.0\\
        \hline
      \end{tabular}
    \end{table}

    \section*{Appendix B: Kernel derivatives}
    This section derives the first and second derivate of the kernel with respect to the coordinates.
    First, we define the signed difference between two representations:
    \begin{align}
    \mathbf{d} = \mathbf{q} - \mathbf{q}^*
    \end{align}
    Derivative of representation wrt. a specific coordinate, $r$, typically the $x$-, $y$-, or $z$-coordinate of an atom in the chemical compound:
    \begin{align}
    \frac{\partial \mathbf{q}}{\partial r} = \left[ \tfrac{\partial q_1}{\partial r} \ \tfrac{\partial q_2}{\partial r} \ \tfrac{\partial q_3}{\partial r} \cdots \tfrac{\partial q_n}{\partial r} \right]^\top
    \end{align}
    Defining a Gaussian kernel:
    \begin{align}
    \mathpzc{k}(\mathbf{q},\mathbf{q}^{*}) = \exp\left(-\frac{\| \mathbf{d}\|^2_2}{2\sigma^2}\right)
    \end{align}
    Defining a vector, $\mathbf{g}$, as the first derivative of the kernel wrt.~$q_i^{*}$:
    \begin{align}
    g_i \triangleq \frac{\partial}{\partial q_i^{*}}\mathpzc{k}(\mathbf{q},\mathbf{q}^{*}) =\frac{d_i}{\sigma^2} \exp\left(-\frac{\| \mathbf{d}\|^2_2}{2\sigma^2}\right)
    \end{align}
    Kernel derivative wrt.~coordinate $r$:
    \begin{align}\label{eq:firstderiv}
    \frac{\partial}{\partial r}\mathpzc{k}(\mathbf{q},\mathbf{q}^{*}) = \mathbf{g} \cdot \left(\frac{\partial \mathbf{q^{*}}}{\partial r}\right)
    \end{align}
    Defining a matrix, $\mathbf{H}$, as the second derivative wrt.~$q_i$ and $q_j^{*}$:
    \begin{align}
    H_{ij} \triangleq \frac{\partial^2}{\partial q_i\partial q_j^{*}}\mathpzc{k}(\mathbf{q},\mathbf{q}^{*}) =\left(\delta_{ij}\frac{1}{\sigma^2}-\frac{d_i d_j}{\sigma^4}\right) \exp\left(-\frac{\| \mathbf{d}\|^2_2}{2\sigma^2}\right)
    \end{align}
    Kernel derivative wrt.~coordinates $r_a$ and $r_b$:
    \begin{align}\label{eq:secondderiv}
    \frac{\partial^2}{\partial r_a\partial r_b^{*}}\mathpzc{k}(\mathbf{q},\mathbf{q}^{*}) = \left( \frac{\partial \mathbf{q}}{\partial r_a}\right)^\top \mathbf{H} \ \left(\frac{\partial \mathbf{q^{*}}}{\partial r^{*}_b}\right)
    \end{align}
    Analytical implementations of these derivatives with the kernel function defined in Eq.~\ref{eq:kernel_screening} are implemented in our \texttt{QML} code.\cite{qmlcode2017}

    \bibliography{refs}

\begin{thebibliography}{69}%
\makeatletter
\providecommand \@ifxundefined [1]{%
 \@ifx{#1\undefined}
}%
\providecommand \@ifnum [1]{%
 \ifnum #1\expandafter \@firstoftwo
 \else \expandafter \@secondoftwo
 \fi
}%
\providecommand \@ifx [1]{%
 \ifx #1\expandafter \@firstoftwo
 \else \expandafter \@secondoftwo
 \fi
}%
\providecommand \natexlab [1]{#1}%
\providecommand \enquote  [1]{``#1''}%
\providecommand \bibnamefont  [1]{#1}%
\providecommand \bibfnamefont [1]{#1}%
\providecommand \citenamefont [1]{#1}%
\providecommand \href@noop [0]{\@secondoftwo}%
\providecommand \href [0]{\begingroup \@sanitize@url \@href}%
\providecommand \@href[1]{\@@startlink{#1}\@@href}%
\providecommand \@@href[1]{\endgroup#1\@@endlink}%
\providecommand \@sanitize@url [0]{\catcode `\\12\catcode `\$12\catcode
  `\&12\catcode `\#12\catcode `\^12\catcode `\_12\catcode `\%12\relax}%
\providecommand \@@startlink[1]{}%
\providecommand \@@endlink[0]{}%
\providecommand \url  [0]{\begingroup\@sanitize@url \@url }%
\providecommand \@url [1]{\endgroup\@href {#1}{\urlprefix }}%
\providecommand \urlprefix  [0]{URL }%
\providecommand \Eprint [0]{\href }%
\providecommand \doibase [0]{http://dx.doi.org/}%
\providecommand \selectlanguage [0]{\@gobble}%
\providecommand \bibinfo  [0]{\@secondoftwo}%
\providecommand \bibfield  [0]{\@secondoftwo}%
\providecommand \translation [1]{[#1]}%
\providecommand \BibitemOpen [0]{}%
\providecommand \bibitemStop [0]{}%
\providecommand \bibitemNoStop [0]{.\EOS\space}%
\providecommand \EOS [0]{\spacefactor3000\relax}%
\providecommand \BibitemShut  [1]{\csname bibitem#1\endcsname}%
\let\auto@bib@innerbib\@empty
\bibitem [{\citenamefont {Faber}\ \emph {et~al.}(2018)\citenamefont {Faber},
  \citenamefont {Christensen}, \citenamefont {Huang},\ and\ \citenamefont {von
  Lilienfeld}}]{faber2017alchemical}%
  \BibitemOpen
  \bibfield  {author} {\bibinfo {author} {\bibfnamefont {F.~A.}\ \bibnamefont
  {Faber}}, \bibinfo {author} {\bibfnamefont {A.~S.}\ \bibnamefont
  {Christensen}}, \bibinfo {author} {\bibfnamefont {B.}~\bibnamefont {Huang}},
  \ and\ \bibinfo {author} {\bibfnamefont {O.~A.}\ \bibnamefont {von
  Lilienfeld}},\ }\bibfield  {title} {\enquote {\bibinfo {title} {Alchemical
  and structural distribution based representation for universal quantum
  machine learning},}\ }\href {\doibase 10.1063/1.5020710} {\bibfield
  {journal} {\bibinfo  {journal} {J. Chem. Phys.}\ }\textbf {\bibinfo {volume}
  {148}},\ \bibinfo {pages} {241717} (\bibinfo {year} {2018})}\BibitemShut
  {NoStop}%
\bibitem [{\citenamefont {Bart\'{o}k}\ and\ \citenamefont
  {Cs\'{a}nyi}(2015)}]{GAPtutorial}%
  \BibitemOpen
  \bibfield  {author} {\bibinfo {author} {\bibfnamefont {A.~P.}\ \bibnamefont
  {Bart\'{o}k}}\ and\ \bibinfo {author} {\bibfnamefont {G.}~\bibnamefont
  {Cs\'{a}nyi}},\ }\bibfield  {title} {\enquote {\bibinfo {title} {Gaussian
  approximation potentials: A brief tutorial introduction},}\ }\href {\doibase
  10.1002/qua.24927} {\bibfield  {journal} {\bibinfo  {journal} {Int. J.
  Quantum Chem.}\ }\textbf {\bibinfo {volume} {115}},\ \bibinfo {pages}
  {1051--1057} (\bibinfo {year} {2015})}\BibitemShut {NoStop}%
\bibitem [{\citenamefont {Lorenz}, \citenamefont {Gross},\ and\ \citenamefont
  {Scheffler}(2004)}]{Neuralnetworks_Scheffler2004}%
  \BibitemOpen
  \bibfield  {author} {\bibinfo {author} {\bibfnamefont {S.}~\bibnamefont
  {Lorenz}}, \bibinfo {author} {\bibfnamefont {A.}~\bibnamefont {Gross}}, \
  and\ \bibinfo {author} {\bibfnamefont {M.}~\bibnamefont {Scheffler}},\
  }\bibfield  {title} {\enquote {\bibinfo {title} {Representing
  high-dimensional potential-energy surfaces for reactions at surfaces by
  neural networks},}\ }\href@noop {} {\bibfield  {journal} {\bibinfo  {journal}
  {Chem. Phys. Lett.}\ }\textbf {\bibinfo {volume} {395}},\ \bibinfo {pages}
  {210} (\bibinfo {year} {2004})}\BibitemShut {NoStop}%
\bibitem [{\citenamefont {Behler}\ and\ \citenamefont
  {Parrinello}(2007)}]{Neuralnetworks_BehlerParrinello2007}%
  \BibitemOpen
  \bibfield  {author} {\bibinfo {author} {\bibfnamefont {J.}~\bibnamefont
  {Behler}}\ and\ \bibinfo {author} {\bibfnamefont {M.}~\bibnamefont
  {Parrinello}},\ }\bibfield  {title} {\enquote {\bibinfo {title} {Generalized
  neural-network representation of high-dimensional potential-energy
  surfaces},}\ }\href@noop {} {\bibfield  {journal} {\bibinfo  {journal} {Phys.
  Rev. Lett.}\ }\textbf {\bibinfo {volume} {98}},\ \bibinfo {pages} {146401}
  (\bibinfo {year} {2007})}\BibitemShut {NoStop}%
\bibitem [{\citenamefont {Behler}(2016)}]{BehlerPerspective2016}%
  \BibitemOpen
  \bibfield  {author} {\bibinfo {author} {\bibfnamefont {J.}~\bibnamefont
  {Behler}},\ }\bibfield  {title} {\enquote {\bibinfo {title} {Perspective:
  Machine learning potentials for atomistic simulations},}\ }\href {\doibase
  10.1063/1.4966192} {\bibfield  {journal} {\bibinfo  {journal} {J. Chem.
  Phys.}\ }\textbf {\bibinfo {volume} {145}},\ \bibinfo {pages} {170901}
  (\bibinfo {year} {2016})}\BibitemShut {NoStop}%
\bibitem [{\citenamefont {Smith}, \citenamefont {Isayev},\ and\ \citenamefont
  {Roitberg}(2017{\natexlab{a}})}]{ANI_IsayevRoitberg2017}%
  \BibitemOpen
  \bibfield  {author} {\bibinfo {author} {\bibfnamefont {J.~S.}\ \bibnamefont
  {Smith}}, \bibinfo {author} {\bibfnamefont {O.}~\bibnamefont {Isayev}}, \
  and\ \bibinfo {author} {\bibfnamefont {A.~E.}\ \bibnamefont {Roitberg}},\
  }\bibfield  {title} {\enquote {\bibinfo {title} {{ANI-1: An extensible neural
  network potential with DFT accuracy at force field computational cost}},}\
  }\href {\doibase 10.1039/C6SC05720A} {\bibfield  {journal} {\bibinfo
  {journal} {Chem. Sci.}\ }\textbf {\bibinfo {volume} {8}},\ \bibinfo {pages}
  {3192--3203} (\bibinfo {year} {2017}{\natexlab{a}})}\BibitemShut {NoStop}%
\bibitem [{\citenamefont {Smith}, \citenamefont {Isayev},\ and\ \citenamefont
  {Roitberg}(2017{\natexlab{b}})}]{smith2017ani}%
  \BibitemOpen
  \bibfield  {author} {\bibinfo {author} {\bibfnamefont {J.~S.}\ \bibnamefont
  {Smith}}, \bibinfo {author} {\bibfnamefont {O.}~\bibnamefont {Isayev}}, \
  and\ \bibinfo {author} {\bibfnamefont {A.~E.}\ \bibnamefont {Roitberg}},\
  }\bibfield  {title} {\enquote {\bibinfo {title} {{ANI-1, A data set of 20
  million calculated off-equilibrium conformations for organic molecules}},}\
  }\href@noop {} {\bibfield  {journal} {\bibinfo  {journal} {Sci. Data}\
  }\textbf {\bibinfo {volume} {4}},\ \bibinfo {pages} {170193} (\bibinfo {year}
  {2017}{\natexlab{b}})}\BibitemShut {NoStop}%
\bibitem [{\citenamefont {Botu}\ and\ \citenamefont
  {Ramprasad}(2015{\natexlab{a}})}]{Botu2015}%
  \BibitemOpen
  \bibfield  {author} {\bibinfo {author} {\bibfnamefont {V.}~\bibnamefont
  {Botu}}\ and\ \bibinfo {author} {\bibfnamefont {R.}~\bibnamefont
  {Ramprasad}},\ }\bibfield  {title} {\enquote {\bibinfo {title} {Learning
  scheme to predict atomic forces and accelerate materials simulations},}\
  }\href@noop {} {\bibfield  {journal} {\bibinfo  {journal} {Phys. Rev. B}\
  }\textbf {\bibinfo {volume} {92}},\ \bibinfo {pages} {094306} (\bibinfo
  {year} {2015}{\natexlab{a}})}\BibitemShut {NoStop}%
\bibitem [{\citenamefont {Botu}\ and\ \citenamefont
  {Ramprasad}(2015{\natexlab{b}})}]{RampiMLQMMM}%
  \BibitemOpen
  \bibfield  {author} {\bibinfo {author} {\bibfnamefont {V.}~\bibnamefont
  {Botu}}\ and\ \bibinfo {author} {\bibfnamefont {R.}~\bibnamefont
  {Ramprasad}},\ }\bibfield  {title} {\enquote {\bibinfo {title} {Adaptive
  machine learning framework to accelerate ab initio molecular dynamics},}\
  }\href {\doibase 10.1002/qua.24836} {\bibfield  {journal} {\bibinfo
  {journal} {Int. J. Quantum Chem.}\ }\textbf {\bibinfo {volume} {115}},\
  \bibinfo {pages} {1074--1083} (\bibinfo {year}
  {2015}{\natexlab{b}})}\BibitemShut {NoStop}%
\bibitem [{\citenamefont {Botu}\ \emph {et~al.}(2016)\citenamefont {Botu},
  \citenamefont {Batra}, \citenamefont {Chapman},\ and\ \citenamefont
  {Ramprasad}}]{Botu2016}%
  \BibitemOpen
  \bibfield  {author} {\bibinfo {author} {\bibfnamefont {V.}~\bibnamefont
  {Botu}}, \bibinfo {author} {\bibfnamefont {R.}~\bibnamefont {Batra}},
  \bibinfo {author} {\bibfnamefont {J.}~\bibnamefont {Chapman}}, \ and\
  \bibinfo {author} {\bibfnamefont {R.}~\bibnamefont {Ramprasad}},\ }\bibfield
  {title} {\enquote {\bibinfo {title} {Machine learning force fields:
  Construction, validation, and outlook},}\ }\href {\doibase
  10.1021/acs.jpcc.6b10908} {\bibfield  {journal} {\bibinfo  {journal} {J.
  Phys. Chem. C}\ }\textbf {\bibinfo {volume} {121}},\ \bibinfo {pages}
  {511--522} (\bibinfo {year} {2016})}\BibitemShut {NoStop}%
\bibitem [{\citenamefont {Huan}\ \emph {et~al.}(2017)\citenamefont {Huan},
  \citenamefont {Batra}, \citenamefont {Chapman}, \citenamefont {Krishnan},
  \citenamefont {Chen},\ and\ \citenamefont {Ramprasad}}]{Huan2017}%
  \BibitemOpen
  \bibfield  {author} {\bibinfo {author} {\bibfnamefont {T.~D.}\ \bibnamefont
  {Huan}}, \bibinfo {author} {\bibfnamefont {R.}~\bibnamefont {Batra}},
  \bibinfo {author} {\bibfnamefont {J.}~\bibnamefont {Chapman}}, \bibinfo
  {author} {\bibfnamefont {S.}~\bibnamefont {Krishnan}}, \bibinfo {author}
  {\bibfnamefont {L.}~\bibnamefont {Chen}}, \ and\ \bibinfo {author}
  {\bibfnamefont {R.}~\bibnamefont {Ramprasad}},\ }\bibfield  {title} {\enquote
  {\bibinfo {title} {A universal strategy for the creation of machine
  learning-based atomistic force fields},}\ }\href {\doibase
  10.1038/s41524-017-0042-y} {\bibfield  {journal} {\bibinfo  {journal} {Npj
  Comput. Mater.}\ }\textbf {\bibinfo {volume} {3}} (\bibinfo {year} {2017}),\
  10.1038/s41524-017-0042-y}\BibitemShut {NoStop}%
\bibitem [{\citenamefont {Li}, \citenamefont {Kermode},\ and\ \citenamefont
  {De~Vita}(2015)}]{Zhenwei2015}%
  \BibitemOpen
  \bibfield  {author} {\bibinfo {author} {\bibfnamefont {Z.}~\bibnamefont
  {Li}}, \bibinfo {author} {\bibfnamefont {J.~R.}\ \bibnamefont {Kermode}}, \
  and\ \bibinfo {author} {\bibfnamefont {A.}~\bibnamefont {De~Vita}},\
  }\bibfield  {title} {\enquote {\bibinfo {title} {Molecular dynamics with
  on-the-fly machine learning of quantum-mechanical forces},}\ }\href {\doibase
  10.1103/PhysRevLett.114.096405} {\bibfield  {journal} {\bibinfo  {journal}
  {Phys. Rev. Lett.}\ }\textbf {\bibinfo {volume} {114}},\ \bibinfo {pages}
  {096405} (\bibinfo {year} {2015})}\BibitemShut {NoStop}%
\bibitem [{\citenamefont {Gubaev}, \citenamefont {Podryabinkin},\ and\
  \citenamefont {Shapeev}(2018)}]{gubaev2018machine}%
  \BibitemOpen
  \bibfield  {author} {\bibinfo {author} {\bibfnamefont {K.}~\bibnamefont
  {Gubaev}}, \bibinfo {author} {\bibfnamefont {E.~V.}\ \bibnamefont
  {Podryabinkin}}, \ and\ \bibinfo {author} {\bibfnamefont {A.~V.}\
  \bibnamefont {Shapeev}},\ }\bibfield  {title} {\enquote {\bibinfo {title}
  {Machine learning of molecular properties: Locality and active learning},}\
  }\href@noop {} {\bibfield  {journal} {\bibinfo  {journal} {J. Chem. Phys.}\
  }\textbf {\bibinfo {volume} {148}},\ \bibinfo {pages} {241727} (\bibinfo
  {year} {2018})}\BibitemShut {NoStop}%
\bibitem [{\citenamefont {Thompson}\ \emph {et~al.}(2015)\citenamefont
  {Thompson}, \citenamefont {Swiler}, \citenamefont {Trott}, \citenamefont
  {Foiles},\ and\ \citenamefont {Tucker}}]{SNAP_Aidan2015}%
  \BibitemOpen
  \bibfield  {author} {\bibinfo {author} {\bibfnamefont {A.}~\bibnamefont
  {Thompson}}, \bibinfo {author} {\bibfnamefont {L.}~\bibnamefont {Swiler}},
  \bibinfo {author} {\bibfnamefont {C.}~\bibnamefont {Trott}}, \bibinfo
  {author} {\bibfnamefont {S.}~\bibnamefont {Foiles}}, \ and\ \bibinfo {author}
  {\bibfnamefont {G.}~\bibnamefont {Tucker}},\ }\bibfield  {title} {\enquote
  {\bibinfo {title} {Spectral neighbor analysis method for automated generation
  of quantum-accurate interatomic potentials},}\ }\href {\doibase
  http://dx.doi.org/10.1016/j.jcp.2014.12.018} {\bibfield  {journal} {\bibinfo
  {journal} {Journal of Computational Physics}\ }\textbf {\bibinfo {volume}
  {285}},\ \bibinfo {pages} {316 -- 330} (\bibinfo {year} {2015})}\BibitemShut
  {NoStop}%
\bibitem [{\citenamefont {Glielmo}, \citenamefont {Sollich},\ and\
  \citenamefont {De~Vita}(2017)}]{CovariantKernelsSandro2016}%
  \BibitemOpen
  \bibfield  {author} {\bibinfo {author} {\bibfnamefont {A.}~\bibnamefont
  {Glielmo}}, \bibinfo {author} {\bibfnamefont {P.}~\bibnamefont {Sollich}}, \
  and\ \bibinfo {author} {\bibfnamefont {A.}~\bibnamefont {De~Vita}},\
  }\bibfield  {title} {\enquote {\bibinfo {title} {Accurate interatomic force
  fields via machine learning with covariant kernels},}\ }\href@noop {}
  {\bibfield  {journal} {\bibinfo  {journal} {Phys. Rev. B}\ }\textbf {\bibinfo
  {volume} {95}},\ \bibinfo {pages} {214302} (\bibinfo {year}
  {2017})}\BibitemShut {NoStop}%
\bibitem [{\citenamefont {Glielmo}, \citenamefont {Zeni},\ and\ \citenamefont
  {De~Vita}(2018)}]{Glielmo2018}%
  \BibitemOpen
  \bibfield  {author} {\bibinfo {author} {\bibfnamefont {A.}~\bibnamefont
  {Glielmo}}, \bibinfo {author} {\bibfnamefont {C.}~\bibnamefont {Zeni}}, \
  and\ \bibinfo {author} {\bibfnamefont {A.}~\bibnamefont {De~Vita}},\
  }\bibfield  {title} {\enquote {\bibinfo {title} {Efficient nonparametric
  $n$-body force fields from machine learning},}\ }\href {\doibase
  10.1103/PhysRevB.97.184307} {\bibfield  {journal} {\bibinfo  {journal} {Phys.
  Rev. B}\ }\textbf {\bibinfo {volume} {97}},\ \bibinfo {pages} {184307}
  (\bibinfo {year} {2018})}\BibitemShut {NoStop}%
\bibitem [{\citenamefont {Sch{\"u}tt}\ \emph {et~al.}(2018)\citenamefont
  {Sch{\"u}tt}, \citenamefont {Sauceda}, \citenamefont {Kindermans},
  \citenamefont {Tkatchenko},\ and\ \citenamefont
  {M{\"u}ller}}]{schutt2018schnet}%
  \BibitemOpen
  \bibfield  {author} {\bibinfo {author} {\bibfnamefont {K.~T.}\ \bibnamefont
  {Sch{\"u}tt}}, \bibinfo {author} {\bibfnamefont {H.~E.}\ \bibnamefont
  {Sauceda}}, \bibinfo {author} {\bibfnamefont {P.-J.}\ \bibnamefont
  {Kindermans}}, \bibinfo {author} {\bibfnamefont {A.}~\bibnamefont
  {Tkatchenko}}, \ and\ \bibinfo {author} {\bibfnamefont {K.-R.}\ \bibnamefont
  {M{\"u}ller}},\ }\bibfield  {title} {\enquote {\bibinfo {title} {{SchNet--A
  deep learning architecture for molecules and materials}},}\ }\href@noop {}
  {\bibfield  {journal} {\bibinfo  {journal} {J. Chem. Phys.}\ }\textbf
  {\bibinfo {volume} {148}},\ \bibinfo {pages} {241722} (\bibinfo {year}
  {2018})}\BibitemShut {NoStop}%
\bibitem [{\citenamefont {Sch{\"u}tt}\ \emph {et~al.}(2019)\citenamefont
  {Sch{\"u}tt}, \citenamefont {Kessel}, \citenamefont {Gastegger},
  \citenamefont {Nicoli}, \citenamefont {Tkatchenko},\ and\ \citenamefont
  {M{\"u}ller}}]{schutt2019schnetpack}%
  \BibitemOpen
  \bibfield  {author} {\bibinfo {author} {\bibfnamefont {K.~T.}\ \bibnamefont
  {Sch{\"u}tt}}, \bibinfo {author} {\bibfnamefont {P.}~\bibnamefont {Kessel}},
  \bibinfo {author} {\bibfnamefont {M.}~\bibnamefont {Gastegger}}, \bibinfo
  {author} {\bibfnamefont {K.~A.}\ \bibnamefont {Nicoli}}, \bibinfo {author}
  {\bibfnamefont {A.}~\bibnamefont {Tkatchenko}}, \ and\ \bibinfo {author}
  {\bibfnamefont {K.-R.}\ \bibnamefont {M{\"u}ller}},\ }\bibfield  {title}
  {\enquote {\bibinfo {title} {{SchNetPack: A Deep Learning Toolbox For
  Atomistic Systems}},}\ }\href {\doibase 10.1021/acs.jctc.8b00908} {\bibfield
  {journal} {\bibinfo  {journal} {J. Chem. Theory Comput.}\ }\textbf {\bibinfo
  {volume} {15}},\ \bibinfo {pages} {448--455} (\bibinfo {year}
  {2019})}\BibitemShut {NoStop}%
\bibitem [{\citenamefont {Grisafi}\ \emph {et~al.}(2018)\citenamefont
  {Grisafi}, \citenamefont {Wilkins}, \citenamefont {Cs{\'a}nyi},\ and\
  \citenamefont {Ceriotti}}]{grisafi2018symmetry}%
  \BibitemOpen
  \bibfield  {author} {\bibinfo {author} {\bibfnamefont {A.}~\bibnamefont
  {Grisafi}}, \bibinfo {author} {\bibfnamefont {D.~M.}\ \bibnamefont
  {Wilkins}}, \bibinfo {author} {\bibfnamefont {G.}~\bibnamefont {Cs{\'a}nyi}},
  \ and\ \bibinfo {author} {\bibfnamefont {M.}~\bibnamefont {Ceriotti}},\
  }\bibfield  {title} {\enquote {\bibinfo {title} {Symmetry-adapted machine
  learning for tensorial properties of atomistic systems},}\ }\href@noop {}
  {\bibfield  {journal} {\bibinfo  {journal} {Phys. Rev. Lett.}\ }\textbf
  {\bibinfo {volume} {120}},\ \bibinfo {pages} {036002} (\bibinfo {year}
  {2018})}\BibitemShut {NoStop}%
\bibitem [{\citenamefont {Zhang}\ \emph {et~al.}(2018)\citenamefont {Zhang},
  \citenamefont {Han}, \citenamefont {Wang}, \citenamefont {Car},\ and\
  \citenamefont {E}}]{DeepMDZhang2018}%
  \BibitemOpen
  \bibfield  {author} {\bibinfo {author} {\bibfnamefont {L.}~\bibnamefont
  {Zhang}}, \bibinfo {author} {\bibfnamefont {J.}~\bibnamefont {Han}}, \bibinfo
  {author} {\bibfnamefont {H.}~\bibnamefont {Wang}}, \bibinfo {author}
  {\bibfnamefont {R.}~\bibnamefont {Car}}, \ and\ \bibinfo {author}
  {\bibfnamefont {W.}~\bibnamefont {E}},\ }\bibfield  {title} {\enquote
  {\bibinfo {title} {Deep potential molecular dynamics: A scalable model with
  the accuracy of quantum mechanics},}\ }\href {\doibase
  10.1103/PhysRevLett.120.143001} {\bibfield  {journal} {\bibinfo  {journal}
  {Phys. Rev. Lett.}\ }\textbf {\bibinfo {volume} {120}},\ \bibinfo {pages}
  {143001} (\bibinfo {year} {2018})}\BibitemShut {NoStop}%
\bibitem [{\citenamefont {Unke}\ and\ \citenamefont
  {Meuwly}(2019)}]{UnkePhysNet2019}%
  \BibitemOpen
  \bibfield  {author} {\bibinfo {author} {\bibfnamefont {O.~T.}\ \bibnamefont
  {Unke}}\ and\ \bibinfo {author} {\bibfnamefont {M.}~\bibnamefont {Meuwly}},\
  }\bibfield  {title} {\enquote {\bibinfo {title} {{PhysNet: A Neural Network
  for Predicting Energies, Forces, Dipole Moments, and Partial Charges}},}\
  }\href {\doibase 10.1021/acs.jctc.9b00181} {\bibfield  {journal} {\bibinfo
  {journal} {J. Chem. Theory Comput.}\ }\textbf {\bibinfo {volume} {15}},\
  \bibinfo {pages} {3678--3693} (\bibinfo {year} {2019})}\BibitemShut {NoStop}%
\bibitem [{\citenamefont {Christensen}, \citenamefont {Faber},\ and\
  \citenamefont {von Lilienfeld}(2019)}]{christensen2018operators}%
  \BibitemOpen
  \bibfield  {author} {\bibinfo {author} {\bibfnamefont {A.~S.}\ \bibnamefont
  {Christensen}}, \bibinfo {author} {\bibfnamefont {F.~A.}\ \bibnamefont
  {Faber}}, \ and\ \bibinfo {author} {\bibfnamefont {O.~A.}\ \bibnamefont {von
  Lilienfeld}},\ }\bibfield  {title} {\enquote {\bibinfo {title} {Operators in
  quantum machine learning: Response properties in chemical space},}\ }\href
  {\doibase 10.1063/1.5053562} {\bibfield  {journal} {\bibinfo  {journal} {J.
  Chem. Phys.}\ }\textbf {\bibinfo {volume} {150}},\ \bibinfo {pages} {064105}
  (\bibinfo {year} {2019})}\BibitemShut {NoStop}%
\bibitem [{\citenamefont {Rogers}\ and\ \citenamefont
  {Hahn}(2010)}]{rogers2010extended}%
  \BibitemOpen
  \bibfield  {author} {\bibinfo {author} {\bibfnamefont {D.}~\bibnamefont
  {Rogers}}\ and\ \bibinfo {author} {\bibfnamefont {M.}~\bibnamefont {Hahn}},\
  }\bibfield  {title} {\enquote {\bibinfo {title} {Extended-connectivity
  fingerprints},}\ }\href@noop {} {\bibfield  {journal} {\bibinfo  {journal}
  {J. Chem. Inf. Model.}\ }\textbf {\bibinfo {volume} {50}},\ \bibinfo {pages}
  {742--754} (\bibinfo {year} {2010})}\BibitemShut {NoStop}%
\bibitem [{\citenamefont {Rupp}\ \emph {et~al.}(2012)\citenamefont {Rupp},
  \citenamefont {Tkatchenko}, \citenamefont {M\"uller},\ and\ \citenamefont
  {von Lilienfeld}}]{RuppPRL2012}%
  \BibitemOpen
  \bibfield  {author} {\bibinfo {author} {\bibfnamefont {M.}~\bibnamefont
  {Rupp}}, \bibinfo {author} {\bibfnamefont {A.}~\bibnamefont {Tkatchenko}},
  \bibinfo {author} {\bibfnamefont {K.-R.}\ \bibnamefont {M\"uller}}, \ and\
  \bibinfo {author} {\bibfnamefont {O.~A.}\ \bibnamefont {von Lilienfeld}},\
  }\bibfield  {title} {\enquote {\bibinfo {title} {Fast and accurate modeling
  of molecular atomization energies with machine learning},}\ }\href@noop {}
  {\bibfield  {journal} {\bibinfo  {journal} {Phys. Rev. Lett.}\ }\textbf
  {\bibinfo {volume} {108}},\ \bibinfo {pages} {058301} (\bibinfo {year}
  {2012})}\BibitemShut {NoStop}%
\bibitem [{\citenamefont {Hansen}\ \emph {et~al.}(2013)\citenamefont {Hansen},
  \citenamefont {Montavon}, \citenamefont {Biegler}, \citenamefont {Fazli},
  \citenamefont {Rupp}, \citenamefont {Scheffler}, \citenamefont {von
  Lilienfeld}, \citenamefont {Tkatchenko},\ and\ \citenamefont
  {M\"uller}}]{AssessmentMLJCTC2013}%
  \BibitemOpen
  \bibfield  {author} {\bibinfo {author} {\bibfnamefont {K.}~\bibnamefont
  {Hansen}}, \bibinfo {author} {\bibfnamefont {G.}~\bibnamefont {Montavon}},
  \bibinfo {author} {\bibfnamefont {F.}~\bibnamefont {Biegler}}, \bibinfo
  {author} {\bibfnamefont {S.}~\bibnamefont {Fazli}}, \bibinfo {author}
  {\bibfnamefont {M.}~\bibnamefont {Rupp}}, \bibinfo {author} {\bibfnamefont
  {M.}~\bibnamefont {Scheffler}}, \bibinfo {author} {\bibfnamefont {O.~A.}\
  \bibnamefont {von Lilienfeld}}, \bibinfo {author} {\bibfnamefont
  {A.}~\bibnamefont {Tkatchenko}}, \ and\ \bibinfo {author} {\bibfnamefont
  {K.-R.}\ \bibnamefont {M\"uller}},\ }\bibfield  {title} {\enquote {\bibinfo
  {title} {Assessment and validation of machine learning methods for predicting
  molecular atomization energies},}\ }\href {\doibase 10.1021/ct400195d}
  {\bibfield  {journal} {\bibinfo  {journal} {J. Chem. Theory Comput.}\
  }\textbf {\bibinfo {volume} {9}},\ \bibinfo {pages} {3404--3419} (\bibinfo
  {year} {2013})}\BibitemShut {NoStop}%
\bibitem [{\citenamefont {Collins}\ \emph {et~al.}(2018)\citenamefont
  {Collins}, \citenamefont {Gordon}, \citenamefont {von Lilienfeld},\ and\
  \citenamefont {Yaron}}]{constsize2018}%
  \BibitemOpen
  \bibfield  {author} {\bibinfo {author} {\bibfnamefont {C.~R.}\ \bibnamefont
  {Collins}}, \bibinfo {author} {\bibfnamefont {G.~J.}\ \bibnamefont {Gordon}},
  \bibinfo {author} {\bibfnamefont {O.~A.}\ \bibnamefont {von Lilienfeld}}, \
  and\ \bibinfo {author} {\bibfnamefont {D.~J.}\ \bibnamefont {Yaron}},\
  }\bibfield  {title} {\enquote {\bibinfo {title} {Constant size descriptors
  for accurate machine learning models of molecular properties},}\ }\href
  {\doibase 10.1063/1.5020441} {\bibfield  {journal} {\bibinfo  {journal} {J.
  Chem. Phys.}\ }\textbf {\bibinfo {volume} {148}},\ \bibinfo {pages} {241718}
  (\bibinfo {year} {2018})}\BibitemShut {NoStop}%
\bibitem [{\citenamefont {von Lilienfeld}\ \emph {et~al.}(2015)\citenamefont
  {von Lilienfeld}, \citenamefont {Ramakrishnan}, \citenamefont {Rupp},\ and\
  \citenamefont {Knoll}}]{FourierDescriptor}%
  \BibitemOpen
  \bibfield  {author} {\bibinfo {author} {\bibfnamefont {O.~A.}\ \bibnamefont
  {von Lilienfeld}}, \bibinfo {author} {\bibfnamefont {R.}~\bibnamefont
  {Ramakrishnan}}, \bibinfo {author} {\bibfnamefont {M.}~\bibnamefont {Rupp}},
  \ and\ \bibinfo {author} {\bibfnamefont {A.}~\bibnamefont {Knoll}},\
  }\bibfield  {title} {\enquote {\bibinfo {title} {Fourier series of atomic
  radial distribution functions: A molecular fingerprint for machine learning
  models of quantum chemical properties},}\ }\href {\doibase 10.1002/qua.24912}
  {\bibfield  {journal} {\bibinfo  {journal} {Int. J. Quantum Chem.}\ }\textbf
  {\bibinfo {volume} {115}},\ \bibinfo {pages} {1084--1093} (\bibinfo {year}
  {2015})}\BibitemShut {NoStop}%
\bibitem [{\citenamefont {Chmiela}\ \emph {et~al.}(2017)\citenamefont
  {Chmiela}, \citenamefont {Tkatchenko}, \citenamefont {Sauceda}, \citenamefont
  {Poltavsky}, \citenamefont {Sch{\"u}tt},\ and\ \citenamefont
  {M{\"u}ller}}]{chmiela2017machine}%
  \BibitemOpen
  \bibfield  {author} {\bibinfo {author} {\bibfnamefont {S.}~\bibnamefont
  {Chmiela}}, \bibinfo {author} {\bibfnamefont {A.}~\bibnamefont {Tkatchenko}},
  \bibinfo {author} {\bibfnamefont {H.~E.}\ \bibnamefont {Sauceda}}, \bibinfo
  {author} {\bibfnamefont {I.}~\bibnamefont {Poltavsky}}, \bibinfo {author}
  {\bibfnamefont {K.~T.}\ \bibnamefont {Sch{\"u}tt}}, \ and\ \bibinfo {author}
  {\bibfnamefont {K.-R.}\ \bibnamefont {M{\"u}ller}},\ }\bibfield  {title}
  {\enquote {\bibinfo {title} {Machine learning of accurate energy-conserving
  molecular force fields},}\ }\href@noop {} {\bibfield  {journal} {\bibinfo
  {journal} {Sci. Adv.}\ }\textbf {\bibinfo {volume} {3}},\ \bibinfo {pages}
  {e1603015} (\bibinfo {year} {2017})}\BibitemShut {NoStop}%
\bibitem [{\citenamefont {Chmiela}\ \emph {et~al.}(2018)\citenamefont
  {Chmiela}, \citenamefont {Sauceda}, \citenamefont {M{\"u}ller},\ and\
  \citenamefont {Tkatchenko}}]{Chmiela2019sGDML}%
  \BibitemOpen
  \bibfield  {author} {\bibinfo {author} {\bibfnamefont {S.}~\bibnamefont
  {Chmiela}}, \bibinfo {author} {\bibfnamefont {H.~E.}\ \bibnamefont
  {Sauceda}}, \bibinfo {author} {\bibfnamefont {K.-R.}\ \bibnamefont
  {M{\"u}ller}}, \ and\ \bibinfo {author} {\bibfnamefont {A.}~\bibnamefont
  {Tkatchenko}},\ }\bibfield  {title} {\enquote {\bibinfo {title} {Towards
  exact molecular dynamics simulations with machine-learned force fields},}\
  }\href@noop {} {\bibfield  {journal} {\bibinfo  {journal} {Nat. Commun.}\
  }\textbf {\bibinfo {volume} {9}},\ \bibinfo {pages} {3887} (\bibinfo {year}
  {2018})}\BibitemShut {NoStop}%
\bibitem [{\citenamefont {Behler}(2011)}]{Neuralnetworks_Behler2011}%
  \BibitemOpen
  \bibfield  {author} {\bibinfo {author} {\bibfnamefont {J.}~\bibnamefont
  {Behler}},\ }\bibfield  {title} {\enquote {\bibinfo {title} {Atom-centered
  symmetry functions for constructing high-dimensional neural networks
  potentials},}\ }\href@noop {} {\bibfield  {journal} {\bibinfo  {journal} {J.
  Chem. Phys.}\ }\textbf {\bibinfo {volume} {134}},\ \bibinfo {pages} {074106}
  (\bibinfo {year} {2011})}\BibitemShut {NoStop}%
\bibitem [{\citenamefont {Gastegger}\ \emph {et~al.}(2018)\citenamefont
  {Gastegger}, \citenamefont {Schwiedrzik}, \citenamefont {Bittermann},
  \citenamefont {Berzsenyi},\ and\ \citenamefont
  {Marquetand}}]{Gastegger2018wacsf}%
  \BibitemOpen
  \bibfield  {author} {\bibinfo {author} {\bibfnamefont {M.}~\bibnamefont
  {Gastegger}}, \bibinfo {author} {\bibfnamefont {L.}~\bibnamefont
  {Schwiedrzik}}, \bibinfo {author} {\bibfnamefont {M.}~\bibnamefont
  {Bittermann}}, \bibinfo {author} {\bibfnamefont {F.}~\bibnamefont
  {Berzsenyi}}, \ and\ \bibinfo {author} {\bibfnamefont {P.}~\bibnamefont
  {Marquetand}},\ }\bibfield  {title} {\enquote {\bibinfo {title}
  {{wACSF—Weighted atom-centered symmetry functions as descriptors in machine
  learning potentials}},}\ }\href {\doibase 10.1063/1.5019667} {\bibfield
  {journal} {\bibinfo  {journal} {J. Chem. Phys.}\ }\textbf {\bibinfo {volume}
  {148}},\ \bibinfo {pages} {241709} (\bibinfo {year} {2018})}\BibitemShut
  {NoStop}%
\bibitem [{\citenamefont {Caro}(2019)}]{SOAPDiscretized2019}%
  \BibitemOpen
  \bibfield  {author} {\bibinfo {author} {\bibfnamefont {M.~A.}\ \bibnamefont
  {Caro}},\ }\bibfield  {title} {\enquote {\bibinfo {title} {Optimizing
  many-body atomic descriptors for enhanced computational performance of
  machine learning based interatomic potentials},}\ }\href {\doibase
  10.1103/PhysRevB.100.024112} {\bibfield  {journal} {\bibinfo  {journal}
  {Phys. Rev. B}\ }\textbf {\bibinfo {volume} {100}},\ \bibinfo {pages}
  {024112} (\bibinfo {year} {2019})}\BibitemShut {NoStop}%
\bibitem [{\citenamefont {Faber}\ \emph {et~al.}(2017)\citenamefont {Faber},
  \citenamefont {Hutchison}, \citenamefont {Huang}, \citenamefont {Gilmer},
  \citenamefont {Schoenholz}, \citenamefont {Dahl}, \citenamefont {Vinyals},
  \citenamefont {Kearnes}, \citenamefont {Riley},\ and\ \citenamefont {von
  Lilienfeld}}]{googlePaper2017}%
  \BibitemOpen
  \bibfield  {author} {\bibinfo {author} {\bibfnamefont {F.~A.}\ \bibnamefont
  {Faber}}, \bibinfo {author} {\bibfnamefont {L.}~\bibnamefont {Hutchison}},
  \bibinfo {author} {\bibfnamefont {B.}~\bibnamefont {Huang}}, \bibinfo
  {author} {\bibfnamefont {J.}~\bibnamefont {Gilmer}}, \bibinfo {author}
  {\bibfnamefont {S.~S.}\ \bibnamefont {Schoenholz}}, \bibinfo {author}
  {\bibfnamefont {G.~E.}\ \bibnamefont {Dahl}}, \bibinfo {author}
  {\bibfnamefont {O.}~\bibnamefont {Vinyals}}, \bibinfo {author} {\bibfnamefont
  {S.}~\bibnamefont {Kearnes}}, \bibinfo {author} {\bibfnamefont {P.~F.}\
  \bibnamefont {Riley}}, \ and\ \bibinfo {author} {\bibfnamefont {O.~A.}\
  \bibnamefont {von Lilienfeld}},\ }\bibfield  {title} {\enquote {\bibinfo
  {title} {Prediction errors of molecular machine learning models lower than
  hybrid {DFT} error},}\ }\href@noop {} {\bibfield  {journal} {\bibinfo
  {journal} {J. Chem. Theory Comput.}\ }\textbf {\bibinfo {volume} {13}},\
  \bibinfo {pages} {5255--5264} (\bibinfo {year} {2017})}\BibitemShut {NoStop}%
\bibitem [{\citenamefont {De}\ \emph {et~al.}(2016)\citenamefont {De},
  \citenamefont {Bartok}, \citenamefont {Csanyi},\ and\ \citenamefont
  {Ceriotti}}]{Sandip2016}%
  \BibitemOpen
  \bibfield  {author} {\bibinfo {author} {\bibfnamefont {S.}~\bibnamefont
  {De}}, \bibinfo {author} {\bibfnamefont {A.~P.}\ \bibnamefont {Bartok}},
  \bibinfo {author} {\bibfnamefont {G.}~\bibnamefont {Csanyi}}, \ and\ \bibinfo
  {author} {\bibfnamefont {M.}~\bibnamefont {Ceriotti}},\ }\bibfield  {title}
  {\enquote {\bibinfo {title} {Comparing molecules and solids across structural
  and alchemical space},}\ }\href {\doibase 10.1039/C6CP00415F} {\bibfield
  {journal} {\bibinfo  {journal} {Phys. Chem. Chem. Phys.}\ }\textbf {\bibinfo
  {volume} {18}},\ \bibinfo {pages} {13754--13769} (\bibinfo {year}
  {2016})}\BibitemShut {NoStop}%
\bibitem [{\citenamefont {Bart\'ok}, \citenamefont {Kondor},\ and\
  \citenamefont {Cs\'anyi}(2013)}]{BartokGabor_Descriptors2013}%
  \BibitemOpen
  \bibfield  {author} {\bibinfo {author} {\bibfnamefont {A.~P.}\ \bibnamefont
  {Bart\'ok}}, \bibinfo {author} {\bibfnamefont {R.}~\bibnamefont {Kondor}}, \
  and\ \bibinfo {author} {\bibfnamefont {G.}~\bibnamefont {Cs\'anyi}},\
  }\bibfield  {title} {\enquote {\bibinfo {title} {On representing chemical
  environments},}\ }\href {\doibase 10.1103/PhysRevB.87.184115} {\bibfield
  {journal} {\bibinfo  {journal} {Phys. Rev. B}\ }\textbf {\bibinfo {volume}
  {87}},\ \bibinfo {pages} {184115} (\bibinfo {year} {2013})}\BibitemShut
  {NoStop}%
\bibitem [{\citenamefont {Huang}\ and\ \citenamefont {von
  Lilienfeld}(2016)}]{Bing2016}%
  \BibitemOpen
  \bibfield  {author} {\bibinfo {author} {\bibfnamefont {B.}~\bibnamefont
  {Huang}}\ and\ \bibinfo {author} {\bibfnamefont {O.~A.}\ \bibnamefont {von
  Lilienfeld}},\ }\bibfield  {title} {\enquote {\bibinfo {title}
  {Communication: Understanding molecular representations in machine learning:
  The role of uniqueness and target similarity},}\ }\href
  {http://scitation.aip.org/content/aip/journal/jcp/145/16/10.1063/1.4964627}
  {\bibfield  {journal} {\bibinfo  {journal} {J. Chem. Phys.}\ }\textbf
  {\bibinfo {volume} {145}},\ \bibinfo {eid} {161102} (\bibinfo {year}
  {2016})}\BibitemShut {NoStop}%
\bibitem [{\citenamefont {Pronobis}, \citenamefont {Tkatchenko},\ and\
  \citenamefont {M{\"u}ller}(2018)}]{pronobis2018many}%
  \BibitemOpen
  \bibfield  {author} {\bibinfo {author} {\bibfnamefont {W.}~\bibnamefont
  {Pronobis}}, \bibinfo {author} {\bibfnamefont {A.}~\bibnamefont
  {Tkatchenko}}, \ and\ \bibinfo {author} {\bibfnamefont {K.-R.}\ \bibnamefont
  {M{\"u}ller}},\ }\bibfield  {title} {\enquote {\bibinfo {title} {Many-body
  descriptors for predicting molecular properties with machine learning:
  Analysis of pairwise and three-body interactions in molecules},}\ }\href@noop
  {} {\bibfield  {journal} {\bibinfo  {journal} {J. Chem. Theory Comput.}\ }
  (\bibinfo {year} {2018})}\BibitemShut {NoStop}%
\bibitem [{\citenamefont {Huang}\ and\ \citenamefont {von
  Lilienfeld}(2017)}]{amons2017}%
  \BibitemOpen
  \bibfield  {author} {\bibinfo {author} {\bibfnamefont {B.}~\bibnamefont
  {Huang}}\ and\ \bibinfo {author} {\bibfnamefont {O.~A.}\ \bibnamefont {von
  Lilienfeld}},\ }\bibfield  {title} {\enquote {\bibinfo {title} {The {``DNA''}
  of chemistry: {Scalable} quantum machine learning with ``amons''},}\
  }\href@noop {} {\bibfield  {journal} {\bibinfo  {journal} {arXiv}\ }
  (\bibinfo {year} {2017})},\ \Eprint {http://arxiv.org/abs/arXiv:1707.04146}
  {arXiv:1707.04146} \BibitemShut {NoStop}%
\bibitem [{\citenamefont {Axilrod}\ and\ \citenamefont
  {Teller}(1943)}]{AxilrodTeller}%
  \BibitemOpen
  \bibfield  {author} {\bibinfo {author} {\bibfnamefont {B.~M.}\ \bibnamefont
  {Axilrod}}\ and\ \bibinfo {author} {\bibfnamefont {E.}~\bibnamefont
  {Teller}},\ }\bibfield  {title} {\enquote {\bibinfo {title} {Interaction of
  the van der {Waals} type between three atoms},}\ }\href@noop {} {\bibfield
  {journal} {\bibinfo  {journal} {J. Chem. Phys.}\ }\textbf {\bibinfo {volume}
  {11}},\ \bibinfo {pages} {299} (\bibinfo {year} {1943})}\BibitemShut
  {NoStop}%
\bibitem [{\citenamefont {Muto}(1943)}]{Muto1943}%
  \BibitemOpen
  \bibfield  {author} {\bibinfo {author} {\bibfnamefont {Y.}~\bibnamefont
  {Muto}},\ }\bibfield  {title} {\enquote {\bibinfo {title} {Force between
  nonpolar molecules},}\ }\href@noop {} {\bibfield  {journal} {\bibinfo
  {journal} {J. Phys.-Math. Soc. Japan}\ }\textbf {\bibinfo {volume} {17}},\
  \bibinfo {pages} {629} (\bibinfo {year} {1943})}\BibitemShut {NoStop}%
\bibitem [{\citenamefont {Rasmussen}\ and\ \citenamefont
  {Williams}(2006)}]{RasmussenWilliams}%
  \BibitemOpen
  \bibfield  {author} {\bibinfo {author} {\bibfnamefont {C.~E.}\ \bibnamefont
  {Rasmussen}}\ and\ \bibinfo {author} {\bibfnamefont {C.~K.~I.}\ \bibnamefont
  {Williams}},\ }\href@noop {} {\emph {\bibinfo {title} {Gaussian Processes for
  Machine Learning, {\tt www.GaussianProcess.org}}}}\ (\bibinfo  {publisher}
  {MIT Press},\ \bibinfo {address} {Cambridge},\ \bibinfo {year} {2006})\
  \bibinfo {note} {editor: T. Dietterich}\BibitemShut {NoStop}%
\bibitem [{\citenamefont {Mathias}(2015)}]{sonjamathias}%
  \BibitemOpen
  \bibfield  {author} {\bibinfo {author} {\bibfnamefont {S.}~\bibnamefont
  {Mathias}},\ }\emph {\bibinfo {title} {{A Kernel-Based Learning Method for an
  efficient Approximation of the high-dimensional Born-Oppenheimer Potential
  Energy Surface}}},\ \href@noop {} {Master's thesis},\ \bibinfo  {school}
  {{Mathematisch-Naturwissenschaftliche Fakult\"at derRheinischen
  Friedrich-Wilhelms-Universit\"at Bonn}}, \bibinfo {address} {Germany}
  (\bibinfo {year} {2015}),\ \bibinfo {note}
  {\url{http://wissrech.ins.uni-bonn.de/teaching/master/masterthesis_mathias_revised.pdf};
  accessed July 2019}\BibitemShut {NoStop}%
\bibitem [{\citenamefont {Hansen}\ \emph {et~al.}(2015)\citenamefont {Hansen},
  \citenamefont {Biegler}, \citenamefont {von Lilienfeld}, \citenamefont
  {M\"uller},\ and\ \citenamefont {Tkatchenko}}]{BobPaper}%
  \BibitemOpen
  \bibfield  {author} {\bibinfo {author} {\bibfnamefont {K.}~\bibnamefont
  {Hansen}}, \bibinfo {author} {\bibfnamefont {F.}~\bibnamefont {Biegler}},
  \bibinfo {author} {\bibfnamefont {O.~A.}\ \bibnamefont {von Lilienfeld}},
  \bibinfo {author} {\bibfnamefont {K.-R.}\ \bibnamefont {M\"uller}}, \ and\
  \bibinfo {author} {\bibfnamefont {A.}~\bibnamefont {Tkatchenko}},\ }\bibfield
   {title} {\enquote {\bibinfo {title} {Interaction potentials in molecules and
  non-local information in chemical space},}\ }\href@noop {} {\bibfield
  {journal} {\bibinfo  {journal} {J. Phys. Chem. Lett.}\ }\textbf {\bibinfo
  {volume} {6}},\ \bibinfo {pages} {2326} (\bibinfo {year} {2015})}\BibitemShut
  {NoStop}%
\bibitem [{\citenamefont {Tikhonov}(1977)}]{tikhonov1977}%
  \BibitemOpen
  \bibfield  {author} {\bibinfo {author} {\bibfnamefont {A.-I.~N.}\
  \bibnamefont {Tikhonov}},\ }\href {https://www.xarg.org/ref/a/0470991240/}
  {\emph {\bibinfo {title} {Solutions of Ill Posed Problems (Scripta series in
  mathematics)}}}\ (\bibinfo  {publisher} {Vh Winston},\ \bibinfo {year}
  {1977})\BibitemShut {NoStop}%
\bibitem [{\citenamefont {Rupp}, \citenamefont {Ramakrishnan},\ and\
  \citenamefont {von Lilienfeld}(2015)}]{MLatoms_2015}%
  \BibitemOpen
  \bibfield  {author} {\bibinfo {author} {\bibfnamefont {M.}~\bibnamefont
  {Rupp}}, \bibinfo {author} {\bibfnamefont {R.}~\bibnamefont {Ramakrishnan}},
  \ and\ \bibinfo {author} {\bibfnamefont {O.~A.}\ \bibnamefont {von
  Lilienfeld}},\ }\bibfield  {title} {\enquote {\bibinfo {title} {Machine
  learning for quantum mechanical properties of atoms in molecules},}\
  }\href@noop {} {\bibfield  {journal} {\bibinfo  {journal} {J. Phys. Chem.
  Lett.}\ }\textbf {\bibinfo {volume} {6}},\ \bibinfo {pages} {3309} (\bibinfo
  {year} {2015})}\BibitemShut {NoStop}%
\bibitem [{\citenamefont {Ramakrishnan}\ \emph
  {et~al.}(2014{\natexlab{a}})\citenamefont {Ramakrishnan}, \citenamefont
  {Dral}, \citenamefont {Rupp},\ and\ \citenamefont {von
  Lilienfeld}}]{DataPaper2014}%
  \BibitemOpen
  \bibfield  {author} {\bibinfo {author} {\bibfnamefont {R.}~\bibnamefont
  {Ramakrishnan}}, \bibinfo {author} {\bibfnamefont {P.}~\bibnamefont {Dral}},
  \bibinfo {author} {\bibfnamefont {M.}~\bibnamefont {Rupp}}, \ and\ \bibinfo
  {author} {\bibfnamefont {O.~A.}\ \bibnamefont {von Lilienfeld}},\ }\bibfield
  {title} {\enquote {\bibinfo {title} {Quantum chemistry structures and
  properties of 134 kilo molecules},}\ }\href@noop {} {\bibfield  {journal}
  {\bibinfo  {journal} {Sci. Data}\ }\textbf {\bibinfo {volume} {1}},\ \bibinfo
  {pages} {140022} (\bibinfo {year} {2014}{\natexlab{a}})}\BibitemShut
  {NoStop}%
\bibitem [{\citenamefont {Waskom}\ \emph {et~al.}(2017)\citenamefont {Waskom},
  \citenamefont {Botvinnik}, \citenamefont {O'Kane}, \citenamefont {Hobson},
  \citenamefont {Lukauskas}, \citenamefont {Gemperline}, \citenamefont
  {Augspurger}, \citenamefont {Halchenko}, \citenamefont {Cole}, \citenamefont
  {Warmenhoven}, \citenamefont {de~Ruiter}, \citenamefont {Pye}, \citenamefont
  {Hoyer}, \citenamefont {Vanderplas}, \citenamefont {Villalba}, \citenamefont
  {Kunter}, \citenamefont {Quintero}, \citenamefont {Bachant}, \citenamefont
  {Martin}, \citenamefont {Meyer}, \citenamefont {Miles}, \citenamefont {Ram},
  \citenamefont {Yarkoni}, \citenamefont {Williams}, \citenamefont {Evans},
  \citenamefont {Fitzgerald}, \citenamefont {Brian}, \citenamefont
  {Fonnesbeck}, \citenamefont {Lee},\ and\ \citenamefont {Qalieh}}]{seaborn}%
  \BibitemOpen
  \bibfield  {author} {\bibinfo {author} {\bibfnamefont {M.}~\bibnamefont
  {Waskom}}, \bibinfo {author} {\bibfnamefont {O.}~\bibnamefont {Botvinnik}},
  \bibinfo {author} {\bibfnamefont {D.}~\bibnamefont {O'Kane}}, \bibinfo
  {author} {\bibfnamefont {P.}~\bibnamefont {Hobson}}, \bibinfo {author}
  {\bibfnamefont {S.}~\bibnamefont {Lukauskas}}, \bibinfo {author}
  {\bibfnamefont {D.~C.}\ \bibnamefont {Gemperline}}, \bibinfo {author}
  {\bibfnamefont {T.}~\bibnamefont {Augspurger}}, \bibinfo {author}
  {\bibfnamefont {Y.}~\bibnamefont {Halchenko}}, \bibinfo {author}
  {\bibfnamefont {J.~B.}\ \bibnamefont {Cole}}, \bibinfo {author}
  {\bibfnamefont {J.}~\bibnamefont {Warmenhoven}}, \bibinfo {author}
  {\bibfnamefont {J.}~\bibnamefont {de~Ruiter}}, \bibinfo {author}
  {\bibfnamefont {C.}~\bibnamefont {Pye}}, \bibinfo {author} {\bibfnamefont
  {S.}~\bibnamefont {Hoyer}}, \bibinfo {author} {\bibfnamefont
  {J.}~\bibnamefont {Vanderplas}}, \bibinfo {author} {\bibfnamefont
  {S.}~\bibnamefont {Villalba}}, \bibinfo {author} {\bibfnamefont
  {G.}~\bibnamefont {Kunter}}, \bibinfo {author} {\bibfnamefont
  {E.}~\bibnamefont {Quintero}}, \bibinfo {author} {\bibfnamefont
  {P.}~\bibnamefont {Bachant}}, \bibinfo {author} {\bibfnamefont
  {M.}~\bibnamefont {Martin}}, \bibinfo {author} {\bibfnamefont
  {K.}~\bibnamefont {Meyer}}, \bibinfo {author} {\bibfnamefont
  {A.}~\bibnamefont {Miles}}, \bibinfo {author} {\bibfnamefont
  {Y.}~\bibnamefont {Ram}}, \bibinfo {author} {\bibfnamefont {T.}~\bibnamefont
  {Yarkoni}}, \bibinfo {author} {\bibfnamefont {M.~L.}\ \bibnamefont
  {Williams}}, \bibinfo {author} {\bibfnamefont {C.}~\bibnamefont {Evans}},
  \bibinfo {author} {\bibfnamefont {C.}~\bibnamefont {Fitzgerald}}, \bibinfo
  {author} {\bibnamefont {Brian}}, \bibinfo {author} {\bibfnamefont
  {C.}~\bibnamefont {Fonnesbeck}}, \bibinfo {author} {\bibfnamefont
  {A.}~\bibnamefont {Lee}}, \ and\ \bibinfo {author} {\bibfnamefont
  {A.}~\bibnamefont {Qalieh}},\ }\href {\doibase 10.5281/zenodo.883859}
  {\enquote {\bibinfo {title} {mwaskom/seaborn: v0.8.1 (september 2017)},}\ }
  (\bibinfo {year} {2017})\BibitemShut {NoStop}%
\bibitem [{\citenamefont {Montavon}\ \emph {et~al.}(2013)\citenamefont
  {Montavon}, \citenamefont {Rupp}, \citenamefont {Gobre}, \citenamefont
  {Vazquez-Mayagoitia}, \citenamefont {Hansen}, \citenamefont {Tkatchenko},
  \citenamefont {M\"uller},\ and\ \citenamefont {von
  Lilienfeld}}]{Montavon2013}%
  \BibitemOpen
  \bibfield  {author} {\bibinfo {author} {\bibfnamefont {G.}~\bibnamefont
  {Montavon}}, \bibinfo {author} {\bibfnamefont {M.}~\bibnamefont {Rupp}},
  \bibinfo {author} {\bibfnamefont {V.}~\bibnamefont {Gobre}}, \bibinfo
  {author} {\bibfnamefont {A.}~\bibnamefont {Vazquez-Mayagoitia}}, \bibinfo
  {author} {\bibfnamefont {K.}~\bibnamefont {Hansen}}, \bibinfo {author}
  {\bibfnamefont {A.}~\bibnamefont {Tkatchenko}}, \bibinfo {author}
  {\bibfnamefont {K.-R.}\ \bibnamefont {M\"uller}}, \ and\ \bibinfo {author}
  {\bibfnamefont {O.~A.}\ \bibnamefont {von Lilienfeld}},\ }\bibfield  {title}
  {\enquote {\bibinfo {title} {Machine learning of molecular electronic
  properties in chemical compound space},}\ }\href
  {http://stacks.iop.org/1367-2630/15/i=9/a=095003} {\bibfield  {journal}
  {\bibinfo  {journal} {New J. Phys.}\ }\textbf {\bibinfo {volume} {15}},\
  \bibinfo {pages} {095003} (\bibinfo {year} {2013})}\BibitemShut {NoStop}%
\bibitem [{\citenamefont {Cheng}\ \emph {et~al.}(2019)\citenamefont {Cheng},
  \citenamefont {Welborn}, \citenamefont {Christensen},\ and\ \citenamefont
  {Miller}}]{mobml2}%
  \BibitemOpen
  \bibfield  {author} {\bibinfo {author} {\bibfnamefont {L.}~\bibnamefont
  {Cheng}}, \bibinfo {author} {\bibfnamefont {M.}~\bibnamefont {Welborn}},
  \bibinfo {author} {\bibfnamefont {A.~S.}\ \bibnamefont {Christensen}}, \ and\
  \bibinfo {author} {\bibfnamefont {T.~F.}\ \bibnamefont {Miller}},\ }\bibfield
   {title} {\enquote {\bibinfo {title} {A universal density matrix functional
  from molecular orbital-based machine learning: Transferability across organic
  molecules},}\ }\href {\doibase 10.1063/1.5088393} {\bibfield  {journal}
  {\bibinfo  {journal} {J. Chem. Phys.}\ }\textbf {\bibinfo {volume} {150}},\
  \bibinfo {pages} {131103} (\bibinfo {year} {2019})}\BibitemShut {NoStop}%
\bibitem [{\citenamefont {Blum}\ and\ \citenamefont
  {Reymond}(2009)}]{ReymondChemicalUniverse3}%
  \BibitemOpen
  \bibfield  {author} {\bibinfo {author} {\bibfnamefont {L.~C.}\ \bibnamefont
  {Blum}}\ and\ \bibinfo {author} {\bibfnamefont {J.-L.}\ \bibnamefont
  {Reymond}},\ }\bibfield  {title} {\enquote {\bibinfo {title} {970 million
  druglike small molecules for virtual screening in the chemical universe
  database {GDB-13}},}\ }\href@noop {} {\bibfield  {journal} {\bibinfo
  {journal} {J. Am. Chem. Soc.}\ }\textbf {\bibinfo {volume} {131}},\ \bibinfo
  {pages} {8732} (\bibinfo {year} {2009})}\BibitemShut {NoStop}%
\bibitem [{\citenamefont {Welborn}, \citenamefont {Cheng},\ and\ \citenamefont
  {Miller}(2018)}]{mobml1}%
  \BibitemOpen
  \bibfield  {author} {\bibinfo {author} {\bibfnamefont {M.}~\bibnamefont
  {Welborn}}, \bibinfo {author} {\bibfnamefont {L.}~\bibnamefont {Cheng}}, \
  and\ \bibinfo {author} {\bibfnamefont {T.~F.}\ \bibnamefont {Miller}},\
  }\bibfield  {title} {\enquote {\bibinfo {title} {Transferability in machine
  learning for electronic structure via the molecular orbital basis},}\ }\href
  {\doibase 10.1021/acs.jctc.8b00636} {\bibfield  {journal} {\bibinfo
  {journal} {J. Chem. Theory Comput.}\ }\textbf {\bibinfo {volume} {14}},\
  \bibinfo {pages} {4772--4779} (\bibinfo {year} {2018})},\ \bibinfo {note}
  {pMID: 30040892}\BibitemShut {NoStop}%
\bibitem [{\citenamefont {Lejaeghere}\ \emph {et~al.}(2016)\citenamefont
  {Lejaeghere}, \citenamefont {Bihlmayer}, \citenamefont {Bj{\"o}rkman},
  \citenamefont {Blaha}, \citenamefont {Bl{\"u}gel}, \citenamefont {Blum},
  \citenamefont {Caliste}, \citenamefont {Castelli}, \citenamefont {Clark},
  \citenamefont {Dal~Corso}, \citenamefont {de~Gironcoli}, \citenamefont
  {Deutsch}, \citenamefont {Dewhurst}, \citenamefont {Di~Marco}, \citenamefont
  {Draxl}, \citenamefont {Du{\l}ak}, \citenamefont {Eriksson}, \citenamefont
  {Flores-Livas}, \citenamefont {Garrity}, \citenamefont {Genovese},
  \citenamefont {Giannozzi}, \citenamefont {Giantomassi}, \citenamefont
  {Goedecker}, \citenamefont {Gonze}, \citenamefont {Gr{\r a}n{\"a}s},
  \citenamefont {Gross}, \citenamefont {Gulans}, \citenamefont {Gygi},
  \citenamefont {Hamann}, \citenamefont {Hasnip}, \citenamefont {Holzwarth},
  \citenamefont {Iu{\c s}an}, \citenamefont {Jochym}, \citenamefont {Jollet},
  \citenamefont {Jones}, \citenamefont {Kresse}, \citenamefont {Koepernik},
  \citenamefont {K{\"u}{\c c}{\"u}kbenli}, \citenamefont {Kvashnin},
  \citenamefont {Locht}, \citenamefont {Lubeck}, \citenamefont {Marsman},
  \citenamefont {Marzari}, \citenamefont {Nitzsche}, \citenamefont
  {Nordstr{\"o}m}, \citenamefont {Ozaki}, \citenamefont {Paulatto},
  \citenamefont {Pickard}, \citenamefont {Poelmans}, \citenamefont {Probert},
  \citenamefont {Refson}, \citenamefont {Richter}, \citenamefont {Rignanese},
  \citenamefont {Saha}, \citenamefont {Scheffler}, \citenamefont {Schlipf},
  \citenamefont {Schwarz}, \citenamefont {Sharma}, \citenamefont {Tavazza},
  \citenamefont {Thunstr{\"o}m}, \citenamefont {Tkatchenko}, \citenamefont
  {Torrent}, \citenamefont {Vanderbilt}, \citenamefont {van Setten},
  \citenamefont {Van~Speybroeck}, \citenamefont {Wills}, \citenamefont {Yates},
  \citenamefont {Zhang},\ and\ \citenamefont {Cottenier}}]{Lejaeghereaad3000}%
  \BibitemOpen
  \bibfield  {author} {\bibinfo {author} {\bibfnamefont {K.}~\bibnamefont
  {Lejaeghere}}, \bibinfo {author} {\bibfnamefont {G.}~\bibnamefont
  {Bihlmayer}}, \bibinfo {author} {\bibfnamefont {T.}~\bibnamefont
  {Bj{\"o}rkman}}, \bibinfo {author} {\bibfnamefont {P.}~\bibnamefont {Blaha}},
  \bibinfo {author} {\bibfnamefont {S.}~\bibnamefont {Bl{\"u}gel}}, \bibinfo
  {author} {\bibfnamefont {V.}~\bibnamefont {Blum}}, \bibinfo {author}
  {\bibfnamefont {D.}~\bibnamefont {Caliste}}, \bibinfo {author} {\bibfnamefont
  {I.~E.}\ \bibnamefont {Castelli}}, \bibinfo {author} {\bibfnamefont {S.~J.}\
  \bibnamefont {Clark}}, \bibinfo {author} {\bibfnamefont {A.}~\bibnamefont
  {Dal~Corso}}, \bibinfo {author} {\bibfnamefont {S.}~\bibnamefont
  {de~Gironcoli}}, \bibinfo {author} {\bibfnamefont {T.}~\bibnamefont
  {Deutsch}}, \bibinfo {author} {\bibfnamefont {J.~K.}\ \bibnamefont
  {Dewhurst}}, \bibinfo {author} {\bibfnamefont {I.}~\bibnamefont {Di~Marco}},
  \bibinfo {author} {\bibfnamefont {C.}~\bibnamefont {Draxl}}, \bibinfo
  {author} {\bibfnamefont {M.}~\bibnamefont {Du{\l}ak}}, \bibinfo {author}
  {\bibfnamefont {O.}~\bibnamefont {Eriksson}}, \bibinfo {author}
  {\bibfnamefont {J.~A.}\ \bibnamefont {Flores-Livas}}, \bibinfo {author}
  {\bibfnamefont {K.~F.}\ \bibnamefont {Garrity}}, \bibinfo {author}
  {\bibfnamefont {L.}~\bibnamefont {Genovese}}, \bibinfo {author}
  {\bibfnamefont {P.}~\bibnamefont {Giannozzi}}, \bibinfo {author}
  {\bibfnamefont {M.}~\bibnamefont {Giantomassi}}, \bibinfo {author}
  {\bibfnamefont {S.}~\bibnamefont {Goedecker}}, \bibinfo {author}
  {\bibfnamefont {X.}~\bibnamefont {Gonze}}, \bibinfo {author} {\bibfnamefont
  {O.}~\bibnamefont {Gr{\r a}n{\"a}s}}, \bibinfo {author} {\bibfnamefont
  {E.~K.~U.}\ \bibnamefont {Gross}}, \bibinfo {author} {\bibfnamefont
  {A.}~\bibnamefont {Gulans}}, \bibinfo {author} {\bibfnamefont
  {F.}~\bibnamefont {Gygi}}, \bibinfo {author} {\bibfnamefont {D.~R.}\
  \bibnamefont {Hamann}}, \bibinfo {author} {\bibfnamefont {P.~J.}\
  \bibnamefont {Hasnip}}, \bibinfo {author} {\bibfnamefont {N.~A.~W.}\
  \bibnamefont {Holzwarth}}, \bibinfo {author} {\bibfnamefont {D.}~\bibnamefont
  {Iu{\c s}an}}, \bibinfo {author} {\bibfnamefont {D.~B.}\ \bibnamefont
  {Jochym}}, \bibinfo {author} {\bibfnamefont {F.}~\bibnamefont {Jollet}},
  \bibinfo {author} {\bibfnamefont {D.}~\bibnamefont {Jones}}, \bibinfo
  {author} {\bibfnamefont {G.}~\bibnamefont {Kresse}}, \bibinfo {author}
  {\bibfnamefont {K.}~\bibnamefont {Koepernik}}, \bibinfo {author}
  {\bibfnamefont {E.}~\bibnamefont {K{\"u}{\c c}{\"u}kbenli}}, \bibinfo
  {author} {\bibfnamefont {Y.~O.}\ \bibnamefont {Kvashnin}}, \bibinfo {author}
  {\bibfnamefont {I.~L.~M.}\ \bibnamefont {Locht}}, \bibinfo {author}
  {\bibfnamefont {S.}~\bibnamefont {Lubeck}}, \bibinfo {author} {\bibfnamefont
  {M.}~\bibnamefont {Marsman}}, \bibinfo {author} {\bibfnamefont
  {N.}~\bibnamefont {Marzari}}, \bibinfo {author} {\bibfnamefont
  {U.}~\bibnamefont {Nitzsche}}, \bibinfo {author} {\bibfnamefont
  {L.}~\bibnamefont {Nordstr{\"o}m}}, \bibinfo {author} {\bibfnamefont
  {T.}~\bibnamefont {Ozaki}}, \bibinfo {author} {\bibfnamefont
  {L.}~\bibnamefont {Paulatto}}, \bibinfo {author} {\bibfnamefont {C.~J.}\
  \bibnamefont {Pickard}}, \bibinfo {author} {\bibfnamefont {W.}~\bibnamefont
  {Poelmans}}, \bibinfo {author} {\bibfnamefont {M.~I.~J.}\ \bibnamefont
  {Probert}}, \bibinfo {author} {\bibfnamefont {K.}~\bibnamefont {Refson}},
  \bibinfo {author} {\bibfnamefont {M.}~\bibnamefont {Richter}}, \bibinfo
  {author} {\bibfnamefont {G.-M.}\ \bibnamefont {Rignanese}}, \bibinfo {author}
  {\bibfnamefont {S.}~\bibnamefont {Saha}}, \bibinfo {author} {\bibfnamefont
  {M.}~\bibnamefont {Scheffler}}, \bibinfo {author} {\bibfnamefont
  {M.}~\bibnamefont {Schlipf}}, \bibinfo {author} {\bibfnamefont
  {K.}~\bibnamefont {Schwarz}}, \bibinfo {author} {\bibfnamefont
  {S.}~\bibnamefont {Sharma}}, \bibinfo {author} {\bibfnamefont
  {F.}~\bibnamefont {Tavazza}}, \bibinfo {author} {\bibfnamefont
  {P.}~\bibnamefont {Thunstr{\"o}m}}, \bibinfo {author} {\bibfnamefont
  {A.}~\bibnamefont {Tkatchenko}}, \bibinfo {author} {\bibfnamefont
  {M.}~\bibnamefont {Torrent}}, \bibinfo {author} {\bibfnamefont
  {D.}~\bibnamefont {Vanderbilt}}, \bibinfo {author} {\bibfnamefont {M.~J.}\
  \bibnamefont {van Setten}}, \bibinfo {author} {\bibfnamefont
  {V.}~\bibnamefont {Van~Speybroeck}}, \bibinfo {author} {\bibfnamefont
  {J.~M.}\ \bibnamefont {Wills}}, \bibinfo {author} {\bibfnamefont {J.~R.}\
  \bibnamefont {Yates}}, \bibinfo {author} {\bibfnamefont {G.-X.}\ \bibnamefont
  {Zhang}}, \ and\ \bibinfo {author} {\bibfnamefont {S.}~\bibnamefont
  {Cottenier}},\ }\bibfield  {title} {\enquote {\bibinfo {title}
  {Reproducibility in density functional theory calculations of solids},}\
  }\href {\doibase 10.1126/science.aad3000} {\bibfield  {journal} {\bibinfo
  {journal} {Science}\ }\textbf {\bibinfo {volume} {351}} (\bibinfo {year}
  {2016}),\ 10.1126/science.aad3000}\BibitemShut {NoStop}%
\bibitem [{\citenamefont {Bootsma}\ and\ \citenamefont
  {Wheeler}(2019)}]{bootsma_wheeler_2019}%
  \BibitemOpen
  \bibfield  {author} {\bibinfo {author} {\bibfnamefont {A.~N.}\ \bibnamefont
  {Bootsma}}\ and\ \bibinfo {author} {\bibfnamefont {S.}~\bibnamefont
  {Wheeler}},\ }\bibfield  {title} {\enquote {\bibinfo {title} {Popular
  integration grids can result in large errors in dft-computed free
  energies},}\ }\href {\doibase 10.26434/chemrxiv.8864204.v1} {\bibfield
  {journal} {\bibinfo  {journal} {ChemRxiv}\ } (\bibinfo {year} {2019}),\
  10.26434/chemrxiv.8864204.v1}\BibitemShut {NoStop}%
\bibitem [{\citenamefont {Christensen}\ \emph {et~al.}(2017)\citenamefont
  {Christensen}, \citenamefont {Faber}, \citenamefont {Huang}, \citenamefont
  {Bratholm}, \citenamefont {Tkatchenko}, \citenamefont {M\"uller},\ and\
  \citenamefont {von Lilienfeld}}]{qmlcode2017}%
  \BibitemOpen
  \bibfield  {author} {\bibinfo {author} {\bibfnamefont {A.~S.}\ \bibnamefont
  {Christensen}}, \bibinfo {author} {\bibfnamefont {F.~A.}\ \bibnamefont
  {Faber}}, \bibinfo {author} {\bibfnamefont {B.}~\bibnamefont {Huang}},
  \bibinfo {author} {\bibfnamefont {L.~A.}\ \bibnamefont {Bratholm}}, \bibinfo
  {author} {\bibfnamefont {A.}~\bibnamefont {Tkatchenko}}, \bibinfo {author}
  {\bibfnamefont {K.-R.}\ \bibnamefont {M\"uller}}, \ and\ \bibinfo {author}
  {\bibfnamefont {O.~A.}\ \bibnamefont {von Lilienfeld}},\ }\href {\doibase
  10.5281/zenodo.817332} {\enquote {\bibinfo {title} {Qml: A python toolkit for
  quantum machine learning, https://github.com/qmlcode/qml},}\ } (\bibinfo
  {year} {2017})\BibitemShut {NoStop}%
\bibitem [{\citenamefont {Ruddigkeit}\ \emph {et~al.}(2012)\citenamefont
  {Ruddigkeit}, \citenamefont {van Deursen}, \citenamefont {Blum},\ and\
  \citenamefont {Reymond}}]{GDB17}%
  \BibitemOpen
  \bibfield  {author} {\bibinfo {author} {\bibfnamefont {L.}~\bibnamefont
  {Ruddigkeit}}, \bibinfo {author} {\bibfnamefont {R.}~\bibnamefont {van
  Deursen}}, \bibinfo {author} {\bibfnamefont {L.}~\bibnamefont {Blum}}, \ and\
  \bibinfo {author} {\bibfnamefont {J.-L.}\ \bibnamefont {Reymond}},\
  }\bibfield  {title} {\enquote {\bibinfo {title} {Enumeration of 166 billion
  organic small molecules in the chemical universe database gdb-17},}\
  }\href@noop {} {\bibfield  {journal} {\bibinfo  {journal} {J. Chem. Inf.
  Model.}\ }\textbf {\bibinfo {volume} {52}},\ \bibinfo {pages} {2684}
  (\bibinfo {year} {2012})}\BibitemShut {NoStop}%
\bibitem [{\citenamefont {Ramakrishnan}\ \emph
  {et~al.}(2014{\natexlab{b}})\citenamefont {Ramakrishnan}, \citenamefont
  {Dral}, \citenamefont {Rupp},\ and\ \citenamefont {von
  Lilienfeld}}]{Ramakrishnan2014uncharacterized}%
  \BibitemOpen
  \bibfield  {author} {\bibinfo {author} {\bibfnamefont {R.}~\bibnamefont
  {Ramakrishnan}}, \bibinfo {author} {\bibfnamefont {P.}~\bibnamefont {Dral}},
  \bibinfo {author} {\bibfnamefont {M.}~\bibnamefont {Rupp}}, \ and\ \bibinfo
  {author} {\bibfnamefont {O.~A.}\ \bibnamefont {von Lilienfeld}},\ }\bibfield
  {title} {\enquote {\bibinfo {title} {{Uncharacterized: List of 3054 molecules
  which failed the geometry consistency check}},}\ }\href {\doibase
  10.6084/m9.figshare.978904\_D10} {\bibfield  {journal} {\bibinfo  {journal}
  {FigShare}\ } (\bibinfo {year} {2014}{\natexlab{b}}),\
  10.6084/m9.figshare.978904\_D10}\BibitemShut {NoStop}%
\bibitem [{\citenamefont {Grimme}\ \emph {et~al.}(2015)\citenamefont {Grimme},
  \citenamefont {Brandenburg}, \citenamefont {Bannwarth},\ and\ \citenamefont
  {Hansen}}]{pbeh3c}%
  \BibitemOpen
  \bibfield  {author} {\bibinfo {author} {\bibfnamefont {S.}~\bibnamefont
  {Grimme}}, \bibinfo {author} {\bibfnamefont {J.~G.}\ \bibnamefont
  {Brandenburg}}, \bibinfo {author} {\bibfnamefont {C.}~\bibnamefont
  {Bannwarth}}, \ and\ \bibinfo {author} {\bibfnamefont {A.}~\bibnamefont
  {Hansen}},\ }\bibfield  {title} {\enquote {\bibinfo {title} {Consistent
  structures and interactions by density functional theory with small atomic
  orbital basis sets},}\ }\href {\doibase 10.1063/1.4927476} {\bibfield
  {journal} {\bibinfo  {journal} {J. Chem. Phys.}\ }\textbf {\bibinfo {volume}
  {143}},\ \bibinfo {pages} {054107} (\bibinfo {year} {2015})}\BibitemShut
  {NoStop}%
\bibitem [{\citenamefont {Christensen}, \citenamefont {Faber},\ and\
  \citenamefont {von Lilienfeld}(2018)}]{training_tar_bz}%
  \BibitemOpen
  \bibfield  {author} {\bibinfo {author} {\bibfnamefont {A.~S.}\ \bibnamefont
  {Christensen}}, \bibinfo {author} {\bibfnamefont {F.~A.}\ \bibnamefont
  {Faber}}, \ and\ \bibinfo {author} {\bibfnamefont {O.~A.}\ \bibnamefont {von
  Lilienfeld}},\ }\bibfield  {title} {\enquote {\bibinfo {title}
  {{training\_data.tar.bz2}},}\ }\href {\doibase
  10.6084/m9.figshare.7000280.v1} {\bibfield  {journal} {\bibinfo  {journal}
  {FigShare}\ } (\bibinfo {year} {2018}),\
  10.6084/m9.figshare.7000280.v1}\BibitemShut {NoStop}%
\bibitem [{\citenamefont {Pedregosa}\ \emph {et~al.}(2011)\citenamefont
  {Pedregosa}, \citenamefont {Varoquaux}, \citenamefont {Gramfort},
  \citenamefont {Michel}, \citenamefont {Thirion}, \citenamefont {Grisel},
  \citenamefont {Blondel}, \citenamefont {Prettenhofer}, \citenamefont {Weiss},
  \citenamefont {Dubourg}, \citenamefont {Vanderplas}, \citenamefont {Passos},
  \citenamefont {Cournapeau}, \citenamefont {Brucher}, \citenamefont {Perrot},\
  and\ \citenamefont {Duchesnay}}]{scikit-learn}%
  \BibitemOpen
  \bibfield  {author} {\bibinfo {author} {\bibfnamefont {F.}~\bibnamefont
  {Pedregosa}}, \bibinfo {author} {\bibfnamefont {G.}~\bibnamefont
  {Varoquaux}}, \bibinfo {author} {\bibfnamefont {A.}~\bibnamefont {Gramfort}},
  \bibinfo {author} {\bibfnamefont {V.}~\bibnamefont {Michel}}, \bibinfo
  {author} {\bibfnamefont {B.}~\bibnamefont {Thirion}}, \bibinfo {author}
  {\bibfnamefont {O.}~\bibnamefont {Grisel}}, \bibinfo {author} {\bibfnamefont
  {M.}~\bibnamefont {Blondel}}, \bibinfo {author} {\bibfnamefont
  {P.}~\bibnamefont {Prettenhofer}}, \bibinfo {author} {\bibfnamefont
  {R.}~\bibnamefont {Weiss}}, \bibinfo {author} {\bibfnamefont
  {V.}~\bibnamefont {Dubourg}}, \bibinfo {author} {\bibfnamefont
  {J.}~\bibnamefont {Vanderplas}}, \bibinfo {author} {\bibfnamefont
  {A.}~\bibnamefont {Passos}}, \bibinfo {author} {\bibfnamefont
  {D.}~\bibnamefont {Cournapeau}}, \bibinfo {author} {\bibfnamefont
  {M.}~\bibnamefont {Brucher}}, \bibinfo {author} {\bibfnamefont
  {M.}~\bibnamefont {Perrot}}, \ and\ \bibinfo {author} {\bibfnamefont
  {E.}~\bibnamefont {Duchesnay}},\ }\bibfield  {title} {\enquote {\bibinfo
  {title} {Scikit-learn: Machine learning in {P}ython},}\ }\href@noop {}
  {\bibfield  {journal} {\bibinfo  {journal} {J. Mach. Learn. Res.}\ }\textbf
  {\bibinfo {volume} {12}},\ \bibinfo {pages} {2825--2830} (\bibinfo {year}
  {2011})}\BibitemShut {NoStop}%
\bibitem [{\citenamefont {Cortes}\ \emph {et~al.}(1994)\citenamefont {Cortes},
  \citenamefont {Jackel}, \citenamefont {Solla}, \citenamefont {Vapnik},\ and\
  \citenamefont {Denker}}]{vapnik1994learningcurves}%
  \BibitemOpen
  \bibfield  {author} {\bibinfo {author} {\bibfnamefont {C.}~\bibnamefont
  {Cortes}}, \bibinfo {author} {\bibfnamefont {L.~D.}\ \bibnamefont {Jackel}},
  \bibinfo {author} {\bibfnamefont {S.~A.}\ \bibnamefont {Solla}}, \bibinfo
  {author} {\bibfnamefont {V.}~\bibnamefont {Vapnik}}, \ and\ \bibinfo {author}
  {\bibfnamefont {J.~S.}\ \bibnamefont {Denker}},\ }\bibfield  {title}
  {\enquote {\bibinfo {title} {Learning curves: Asymptotic values and rate of
  convergence},}\ }in\ \href@noop {} {\emph {\bibinfo {booktitle} {Advances in
  Neural Information Processing Systems}}}\ (\bibinfo  {publisher}
  {Morgan-Kaufsmann},\ \bibinfo {year} {1994})\ pp.\ \bibinfo {pages}
  {327--334}\BibitemShut {NoStop}%
\bibitem [{\citenamefont {{M\"uller}}\ \emph {et~al.}(1996)\citenamefont
  {{M\"uller}}, \citenamefont {Finke}, \citenamefont {Murata}, \citenamefont
  {Schulten},\ and\ \citenamefont {Amari}}]{StatError_Muller1996}%
  \BibitemOpen
  \bibfield  {author} {\bibinfo {author} {\bibfnamefont {K.~R.}\ \bibnamefont
  {{M\"uller}}}, \bibinfo {author} {\bibfnamefont {M.}~\bibnamefont {Finke}},
  \bibinfo {author} {\bibfnamefont {N.}~\bibnamefont {Murata}}, \bibinfo
  {author} {\bibfnamefont {K.}~\bibnamefont {Schulten}}, \ and\ \bibinfo
  {author} {\bibfnamefont {S.}~\bibnamefont {Amari}},\ }\bibfield  {title}
  {\enquote {\bibinfo {title} {A numerical study on learning curves in
  stochastic multilayer feedforward networks},}\ }\href@noop {} {\bibfield
  {journal} {\bibinfo  {journal} {Neural Comp.}\ }\textbf {\bibinfo {volume}
  {8}},\ \bibinfo {pages} {1085} (\bibinfo {year} {1996})}\BibitemShut
  {NoStop}%
\bibitem [{\citenamefont {von Lilienfeld}(2018)}]{QMLessayAnatole}%
  \BibitemOpen
  \bibfield  {author} {\bibinfo {author} {\bibfnamefont {O.~A.}\ \bibnamefont
  {von Lilienfeld}},\ }\bibfield  {title} {\enquote {\bibinfo {title} {Quantum
  machine learning in chemical compound space},}\ }\href {\doibase
  10.1002/anie.201709686} {\bibfield  {journal} {\bibinfo  {journal} {Angew.
  Chem. Int. Ed.}\ }\textbf {\bibinfo {volume} {57}},\ \bibinfo {pages} {4164}
  (\bibinfo {year} {2018})}\BibitemShut {NoStop}%
\bibitem [{\citenamefont {Deeks}\ \emph {et~al.}(2019)\citenamefont {Deeks},
  \citenamefont {Walters}, \citenamefont {Hare}, \citenamefont {O'Connor},
  \citenamefont {Mulholland},\ and\ \citenamefont {Glowacki}}]{Deeks2019}%
  \BibitemOpen
  \bibfield  {author} {\bibinfo {author} {\bibfnamefont {H.~M.}\ \bibnamefont
  {Deeks}}, \bibinfo {author} {\bibfnamefont {R.~K.}\ \bibnamefont {Walters}},
  \bibinfo {author} {\bibfnamefont {S.~R.}\ \bibnamefont {Hare}}, \bibinfo
  {author} {\bibfnamefont {M.~B.}\ \bibnamefont {O'Connor}}, \bibinfo {author}
  {\bibfnamefont {A.~J.}\ \bibnamefont {Mulholland}}, \ and\ \bibinfo {author}
  {\bibfnamefont {D.~R.}\ \bibnamefont {Glowacki}},\ }\bibfield  {title}
  {\enquote {\bibinfo {title} {Sampling protein-ligand binding pathways to
  recover crystallographic binding poses using interactive molecular dynamics
  in virtual reality},}\ }\href@noop {} {\  (\bibinfo {year} {2019})},\ \Eprint
  {http://arxiv.org/abs/arXiv:1908.07395} {arXiv:1908.07395} \BibitemShut
  {NoStop}%
\bibitem [{\citenamefont {Amabilino}\ \emph {et~al.}(2019)\citenamefont
  {Amabilino}, \citenamefont {Bratholm}, \citenamefont {Bennie}, \citenamefont
  {Vaucher}, \citenamefont {Reiher},\ and\ \citenamefont
  {Glowacki}}]{Amabilino2019}%
  \BibitemOpen
  \bibfield  {author} {\bibinfo {author} {\bibfnamefont {S.}~\bibnamefont
  {Amabilino}}, \bibinfo {author} {\bibfnamefont {L.~A.}\ \bibnamefont
  {Bratholm}}, \bibinfo {author} {\bibfnamefont {S.~J.}\ \bibnamefont
  {Bennie}}, \bibinfo {author} {\bibfnamefont {A.~C.}\ \bibnamefont {Vaucher}},
  \bibinfo {author} {\bibfnamefont {M.}~\bibnamefont {Reiher}}, \ and\ \bibinfo
  {author} {\bibfnamefont {D.~R.}\ \bibnamefont {Glowacki}},\ }\bibfield
  {title} {\enquote {\bibinfo {title} {Training neural nets to learn reactive
  potential energy surfaces using interactive quantum chemistry in virtual
  reality},}\ }\href {\doibase 10.1021/acs.jpca.9b01006} {\bibfield  {journal}
  {\bibinfo  {journal} {The Journal of Physical Chemistry A}\ }\textbf
  {\bibinfo {volume} {123}},\ \bibinfo {pages} {4486--4499} (\bibinfo {year}
  {2019})}\BibitemShut {NoStop}%
\bibitem [{\citenamefont {O'Connor}\ \emph {et~al.}(2018)\citenamefont
  {O'Connor}, \citenamefont {Deeks}, \citenamefont {Dawn}, \citenamefont
  {Metatla}, \citenamefont {Roudaut}, \citenamefont {Sutton}, \citenamefont
  {Thomas}, \citenamefont {Glowacki}, \citenamefont {Sage}, \citenamefont
  {Tew}, \citenamefont {Wonnacott}, \citenamefont {Bates}, \citenamefont
  {Mulholland},\ and\ \citenamefont {Glowacki}}]{OConnor2019}%
  \BibitemOpen
  \bibfield  {author} {\bibinfo {author} {\bibfnamefont {M.}~\bibnamefont
  {O'Connor}}, \bibinfo {author} {\bibfnamefont {H.~M.}\ \bibnamefont {Deeks}},
  \bibinfo {author} {\bibfnamefont {E.}~\bibnamefont {Dawn}}, \bibinfo {author}
  {\bibfnamefont {O.}~\bibnamefont {Metatla}}, \bibinfo {author} {\bibfnamefont
  {A.}~\bibnamefont {Roudaut}}, \bibinfo {author} {\bibfnamefont
  {M.}~\bibnamefont {Sutton}}, \bibinfo {author} {\bibfnamefont {L.~M.}\
  \bibnamefont {Thomas}}, \bibinfo {author} {\bibfnamefont {B.~R.}\
  \bibnamefont {Glowacki}}, \bibinfo {author} {\bibfnamefont {R.}~\bibnamefont
  {Sage}}, \bibinfo {author} {\bibfnamefont {P.}~\bibnamefont {Tew}}, \bibinfo
  {author} {\bibfnamefont {M.}~\bibnamefont {Wonnacott}}, \bibinfo {author}
  {\bibfnamefont {P.}~\bibnamefont {Bates}}, \bibinfo {author} {\bibfnamefont
  {A.~J.}\ \bibnamefont {Mulholland}}, \ and\ \bibinfo {author} {\bibfnamefont
  {D.~R.}\ \bibnamefont {Glowacki}},\ }\bibfield  {title} {\enquote {\bibinfo
  {title} {Sampling molecular conformations and dynamics in a multiuser virtual
  reality framework},}\ }\href {\doibase 10.1126/sciadv.aat2731} {\bibfield
  {journal} {\bibinfo  {journal} {Sci. Adv.}\ }\textbf {\bibinfo {volume} {4}}
  (\bibinfo {year} {2018}),\ 10.1126/sciadv.aat2731}\BibitemShut {NoStop}%
\bibitem [{\citenamefont {Bennie}\ \emph {et~al.}(2019)\citenamefont {Bennie},
  \citenamefont {Ranaghan}, \citenamefont {Deeks}, \citenamefont {Goldsmith},
  \citenamefont {O'Connor}, \citenamefont {Mulholland},\ and\ \citenamefont
  {Glowacki}}]{Bennie2019}%
  \BibitemOpen
  \bibfield  {author} {\bibinfo {author} {\bibfnamefont {S.~J.}\ \bibnamefont
  {Bennie}}, \bibinfo {author} {\bibfnamefont {K.~E.}\ \bibnamefont
  {Ranaghan}}, \bibinfo {author} {\bibfnamefont {H.}~\bibnamefont {Deeks}},
  \bibinfo {author} {\bibfnamefont {H.~E.}\ \bibnamefont {Goldsmith}}, \bibinfo
  {author} {\bibfnamefont {M.~B.}\ \bibnamefont {O'Connor}}, \bibinfo {author}
  {\bibfnamefont {A.~J.}\ \bibnamefont {Mulholland}}, \ and\ \bibinfo {author}
  {\bibfnamefont {D.~R.}\ \bibnamefont {Glowacki}},\ }\bibfield  {title}
  {\enquote {\bibinfo {title} {Teaching enzyme catalysis using interactive
  molecular dynamics in virtual reality},}\ }\href {\doibase
  10.1021/acs.jchemed.9b00181} {\bibfield  {journal} {\bibinfo  {journal}
  {Journal of Chemical Education}\ } (\bibinfo {year} {2019}),\
  10.1021/acs.jchemed.9b00181}\BibitemShut {NoStop}%
\bibitem [{\citenamefont {Ferrell}\ \emph {et~al.}(2019)\citenamefont
  {Ferrell}, \citenamefont {Campbell}, \citenamefont {McCarthy}, \citenamefont
  {McKay}, \citenamefont {Hensinger}, \citenamefont {Srinivasan}, \citenamefont
  {Zhao}, \citenamefont {Wurthmann}, \citenamefont {Li},\ and\ \citenamefont
  {Schneebeli}}]{Ferrell2019}%
  \BibitemOpen
  \bibfield  {author} {\bibinfo {author} {\bibfnamefont {J.~B.}\ \bibnamefont
  {Ferrell}}, \bibinfo {author} {\bibfnamefont {J.~P.}\ \bibnamefont
  {Campbell}}, \bibinfo {author} {\bibfnamefont {D.~R.}\ \bibnamefont
  {McCarthy}}, \bibinfo {author} {\bibfnamefont {K.~T.}\ \bibnamefont {McKay}},
  \bibinfo {author} {\bibfnamefont {M.}~\bibnamefont {Hensinger}}, \bibinfo
  {author} {\bibfnamefont {R.}~\bibnamefont {Srinivasan}}, \bibinfo {author}
  {\bibfnamefont {X.}~\bibnamefont {Zhao}}, \bibinfo {author} {\bibfnamefont
  {A.}~\bibnamefont {Wurthmann}}, \bibinfo {author} {\bibfnamefont
  {J.}~\bibnamefont {Li}}, \ and\ \bibinfo {author} {\bibfnamefont {S.~T.}\
  \bibnamefont {Schneebeli}},\ }\bibfield  {title} {\enquote {\bibinfo {title}
  {Chemical exploration with virtual reality in organic teaching
  laboratories},}\ }\href {\doibase 10.1021/acs.jchemed.9b00036} {\bibfield
  {journal} {\bibinfo  {journal} {Journal of Chemical Education}\ } (\bibinfo
  {year} {2019}),\ 10.1021/acs.jchemed.9b00036}\BibitemShut {NoStop}%
\bibitem [{\citenamefont {Oliphant}(06  )}]{numpy}%
  \BibitemOpen
  \bibfield  {author} {\bibinfo {author} {\bibfnamefont {T.}~\bibnamefont
  {Oliphant}},\ }\href {http://www.numpy.org/} {\enquote {\bibinfo {title}
  {{NumPy}: A guide to {NumPy}},}\ }\bibinfo {howpublished} {USA: Trelgol
  Publishing} (\bibinfo {year} {2006--}),\ \bibinfo {note} {[Online; accessed
  July 2019]}\BibitemShut {NoStop}%
\bibitem [{\citenamefont {{OpenMP Architecture Review
  Board}}(2008)}]{openmp08}%
  \BibitemOpen
  \bibfield  {author} {\bibinfo {author} {\bibnamefont {{OpenMP Architecture
  Review Board}}},\ }\href {http://www.openmp.org/mp-documents/spec30.pdf}
  {\enquote {\bibinfo {title} {{OpenMP} application program interface version
  3.0},}\ } (\bibinfo {year} {2008})\BibitemShut {NoStop}%
\end{thebibliography}%
    
    \end{document}